\documentclass{ws-procs975x65}
\usepackage{graphicx}

\newcommand{\Mpc}{$h^{-1}$\thinspace Mpc}

\newcommand{\etal}{et al.}
\def\apj{ApJ}
\def\aj{Astron. J.}
 
\def\apjs{ApJS}  

\def\aap{A\&A}

\def\mnras{MNRAS}
\def\azh{AstrZh}
\def\bain{Bull. Astr. Inst. Neth.}
\def\nat{Nature}
\def\araa{Ann. Rev. Astr. Astroph.}

\begin{document}

\title{Two hundred years of galactic studies in Tartu Observatory}

\author {J. Einasto}
\address{Tartu Observatory, EE-61602 T\~oravere, Estonia}

\begin{abstract}
  An overview is provided for 200 years of galactic studies at the  Tartu
  Observatory.  Galactic 
  studies have been one of the main topics of studies in Tartu over the whole
  period of the history of the Observatory, starting from F.G.W. Struve and
  J.H. M\"adler, followed by Ernst \"Opik and Grigori Kuzmin, and continuing
  with the present generation of astronomers. Our goal was to understand
  better the structure, origin and evolution of stars, galaxies and the
  Universe.
\end{abstract}


\bodymatter

\section{Introduction}

Galactic studies have been one of the main topics of the Tartu Observatory over the whole period of its existence. These studies began with the work by F.G.W. Struve on double stars and measurements of stellar parallaxes in early 19th century, when Tartu Observatory was founded in the ''Kaiserliche Universit\"at zu Dorpat''.  Modern era of Galactic studies began about 95 years ago when Ernst \"Opik determined the dynamical density of matter in the disk of the Galaxy in 1915, the distance to the Andromeda Nebula in 1922, and found main principles of stellar structure and evolution in 1938.  His student Grigori Kuzmin developed principles of Galactic modeling and calculated the local density of matter near the Sun, suggesting the absence of local dark matter in large quantities. The present generation of astronomers follows these traditions.

In the following I give an overview of the development of the ideas of the structure of galaxies and the Universe over the whole period of activity of the Tartu Observatory. The period of last 50 years is described in more detail.

\section{Tartu Observatory  1810 - 1960}

\subsection{F. G. W. Struve, stellar distances and the Meridian Arc}

\begin{figure*}[ht]
\centering
\resizebox{.50\columnwidth}{!}{\includegraphics*{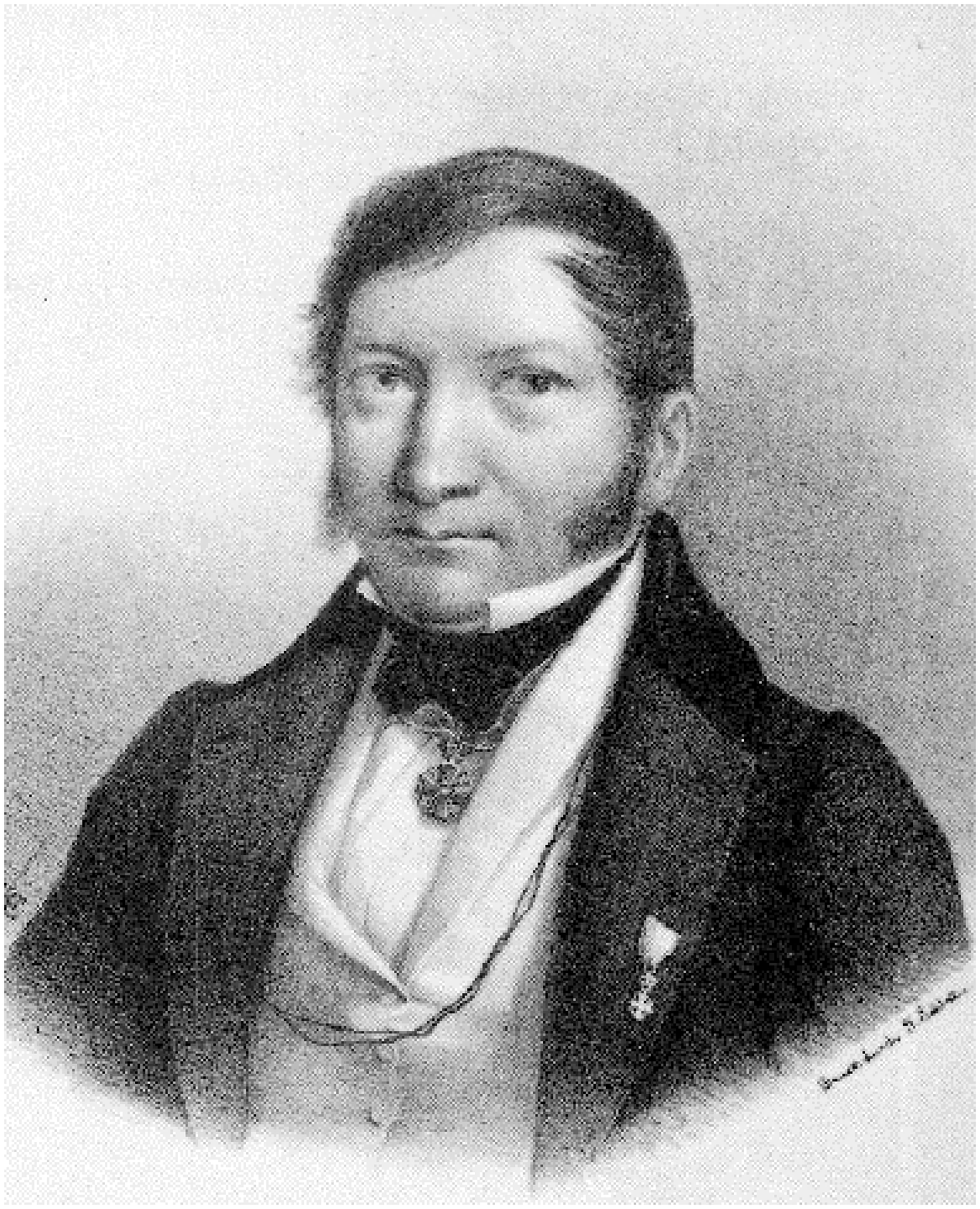}}
\resizebox{.42\columnwidth}{!}{\includegraphics*{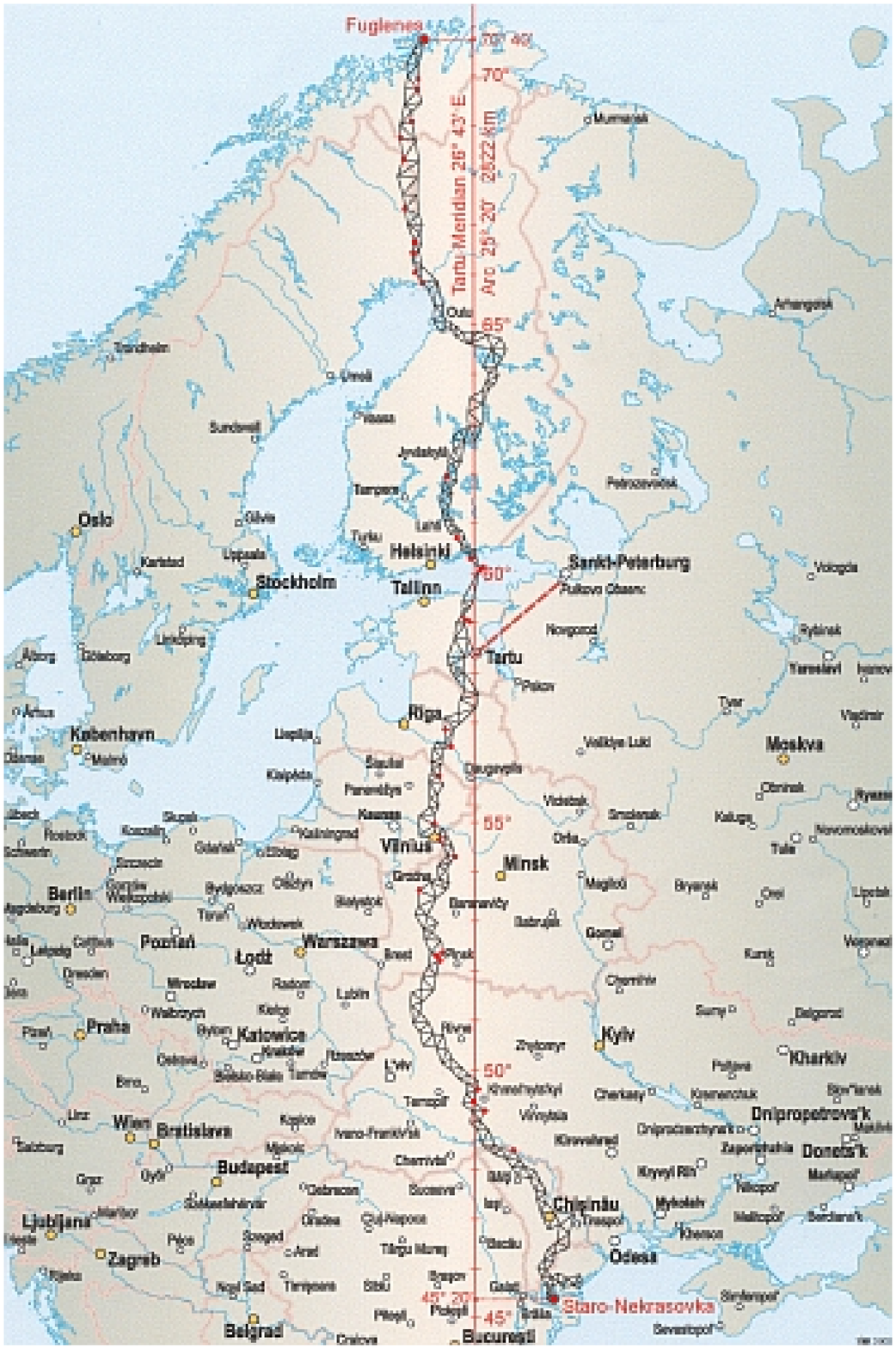}}
\caption{Left: F.G.W Struve. Right: Map of Struve Meridian Arc through Tartu
  Observatory }
\label{fig:struve}
\end{figure*}

 It was a tradition in classical universities to have an astronomical
 observatory. Tartu University Observatory was built in 1810, its astronomical
 instruments were installed by astronomy and mathematics student Friedrich
 Georg Wilhelm Struve (1793 - 1864), who after defending his PhD was nominated
 to professor of astronomy and mathematics and director of the observatory.
 Struve understood well the needs of contemporary astronomy and geodesy.  The
 main goal at this time was to understand the nature of stars and stellar
 systems.  Struve started to measure stellar positions, which was basis for
 better description of the universe.  He soon realised that conventional
 astronomical instruments, available in Tartu, were not sufficient to solve most
 interesting problems. Thus he applied to get support to buy a new larger
 telescope.  His efforts succeeded and in 1825 a 9 inch Fraunhofer refractor
 was installed in Observatory. For about 15 years this was the largest and
 best telescope of this type in the world.

Struve made excellent use of the new telescope.  First he made a
survey of the whole Northern sky and published a catalogue of double
stars detected. A catalogue followed, which contained exact
measurements of positions of double stars\cite{Struve:1837}.  After
repeated observations have been made it is possible to calculate
orbits of double stars and to get information on their masses.  This
is a foundation of the new astronomy -- astrophysics.  As a by-product
of the measurements of double stars he also published his
determination of the distance of a star -- Vega.  With this measurement
it was finally demonstrated that stars are distant suns.  Struve
double star catalogue is used even in present days, his achievements
have found place in astronomy textbooks.

Another important scientific-historic achievement of F.G.W. Struve was
the astronomic-trigonometric measurement of Tartu (Struve) Meridian
Arc (1816-1852). This measurement was made together with Carl
Friedrich Tenner (1783 - 1859). They succeeded in determining the
almost 3000 km long section of the Meridian Arc between the mouth of
Danube and the Arctic Ocean with the accuracy of $\pm 12$~m
\cite{Struve:1860}. The comparison of data of different arc sections
indicated that the length of arc, corresponding to one degree of
latitude, increases towards the pole, i.e.\ the Earth is flattened. The
measurements were used by F.W. Bessel for determination of spheroidal
Earth’s new parameters. Struve geodetic Meridian Arc is included to
the UNESCO World Heritage List.

In 1839 F.G.W. Struve was appointed director of the new Pulkovo
Observatory. The next director of Tartu Observatory Johann
Heinrich M\"adler (1794 - 1874) was interested in the dynamics of the
Milky Way. In his book  ''Centralsonne'' he tried to determine the
center of the Galaxy using proper motions of stars. The accuracy of
data was not sufficient for this task, the method itself is correct.

\subsection{Ernst \"Opik, the nature of galaxies and the evolution of stars}

\begin{figure*}[ht]
\centering
\resizebox{.40\columnwidth}{!}{\includegraphics*{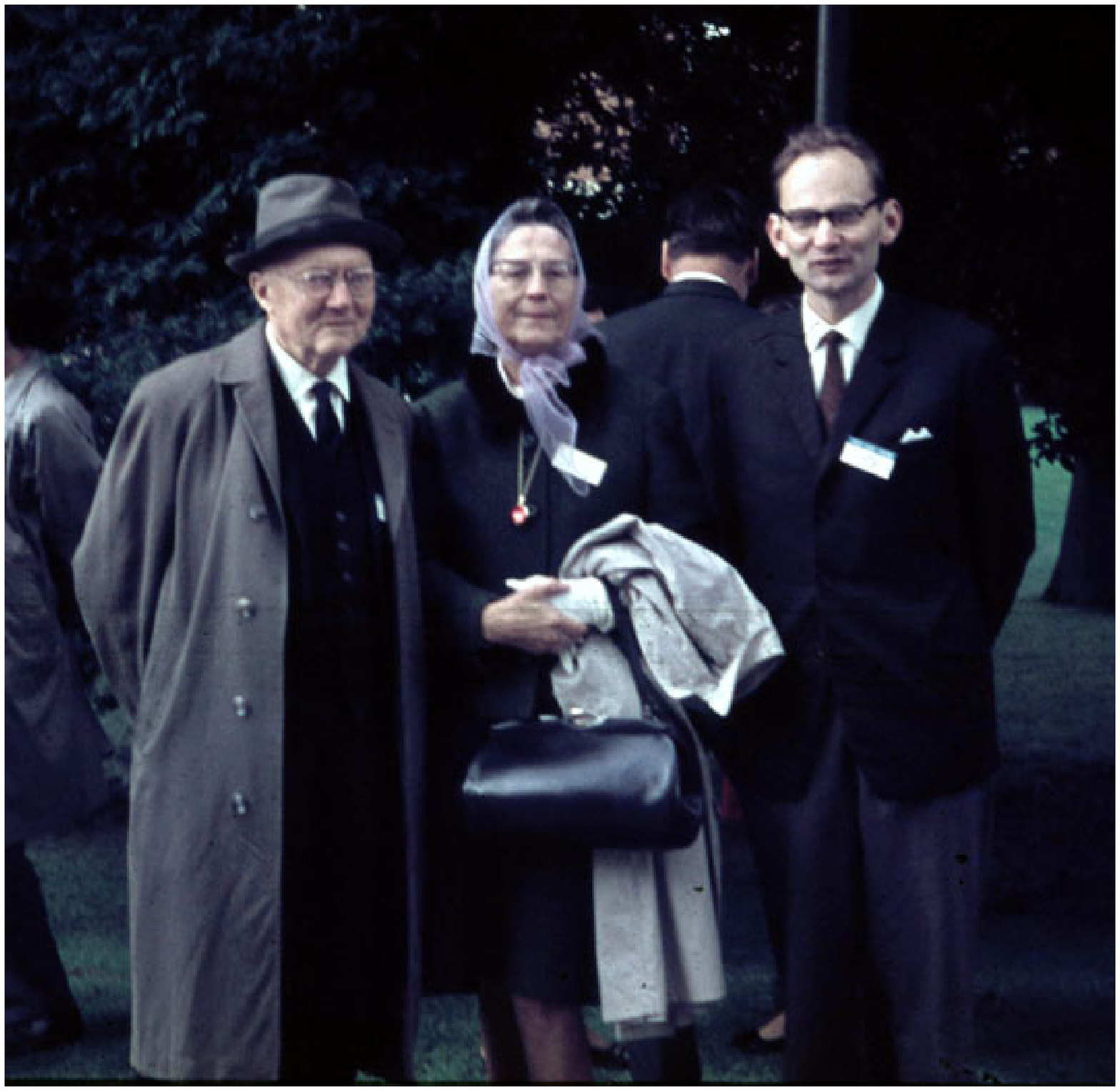}}
\resizebox{.57\columnwidth}{!}{\includegraphics*{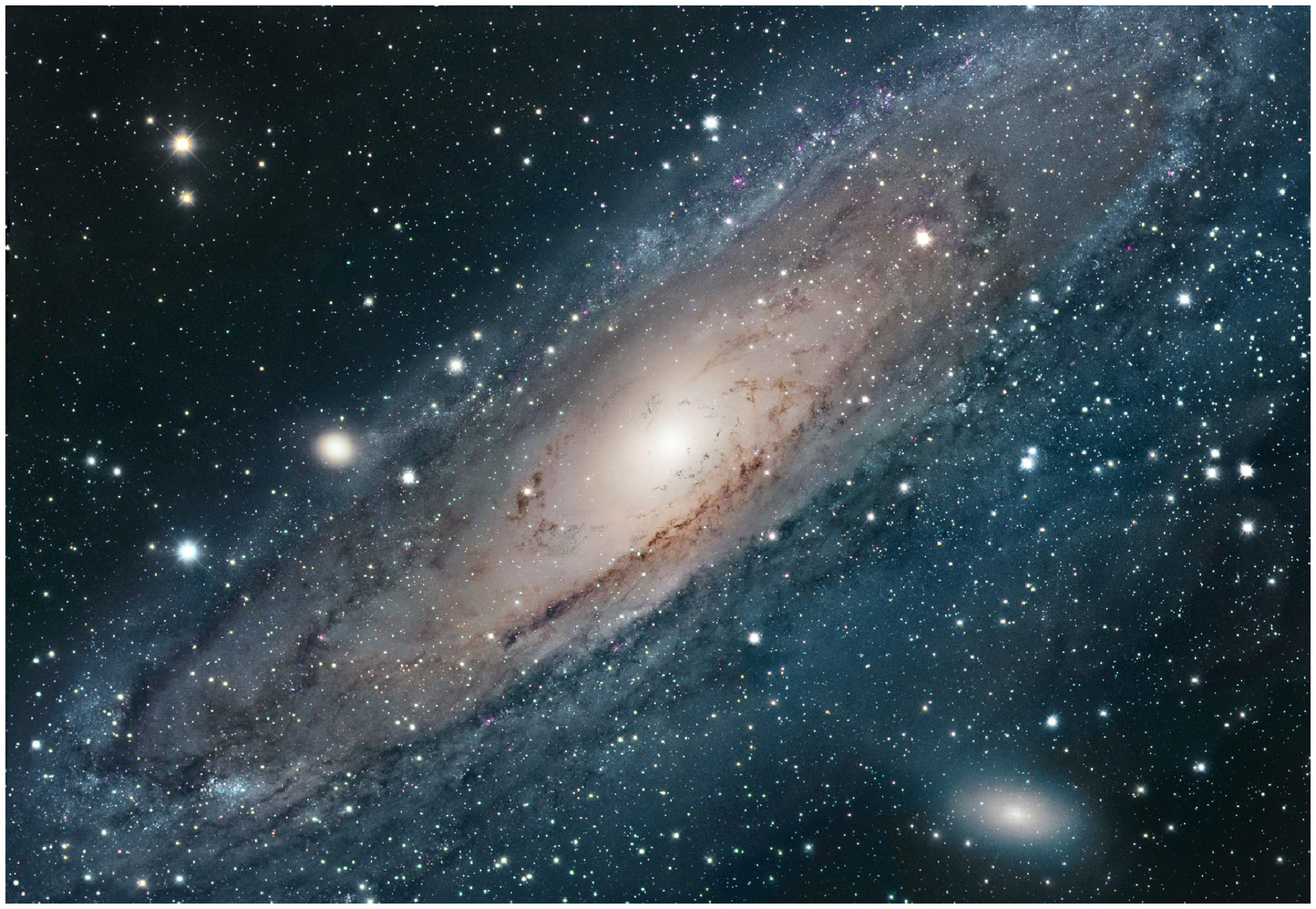}}
\caption{Ernst \"Opik, his wife and author in 1970  at the IAU General
  Assembly in Brighton.  Andromeda nebula M31.  }
\label{fig:opik}
\end{figure*}

The founder of the modern astronomy school in Tartu University was Ernst
\"Opik (1893 - 1985).  He started his astronomical career as student of the
Moscow University, and made his principal discoveries as observator of the
Tartu University Observatory.  One of his first scientific papers was devoted
to the question: What is the density of matter near the plane of the Milky Way
stellar system?  His calculations showed that the gravitating matter density
can be fully explained by observable stars, and that there is no evidence for
hypothetical matter near the symmetry plane of the Galaxy.  This work seems to
be the first attempt to address the dark matter problem\cite{Opik:1915}.

Next \"Opik devoted his attention to spiral nebulae.  Their nature was
not known at this time: Are they gaseous objects within the Milky Way
or distant worlds similar in structure to our Galaxy?  Immediately
when relative velocity measurements of the central part of Andromeda
nebula M31 had been published, \"Opik developed a method how to use this
information to estimate the distance to M31.  His result was 440 kpc (about
1.5 million light years)\cite{Opik:1922}. With this work he solved the
problem of the nature of spiral nebulae, and showed that the universe
is millions of times larger than our Milky Way system.

Another unsolved problem at this time was the source of stellar energy
and the evolution of stars.  Already in early 1920s \"Opik demonstrated, using
very simple physical considerations, that gravitational contraction and
radioactivity cannot be the main sources of stellar energy.   He 
concluded, that some unknown subatomic processes in central regions of
stars must be responsible for stellar energy.  When data on atomic
structure were available, he developed a detailed theory of stellar
evolution, based on nuclear reactions in stellar interiors.  Here
under very high temperature hydrogen burns to helium, and huge amounts
of energy will be released \cite{Opik:1938}.  The efficiency of these
reactions has just been estimated based on atomic theory, thus \"Opik
was able to calculate ages of stars.  He demonstrated, that hot giant
stars are so luminous that their energy sources will be exhausted
within several tens  millions  years.  Their presence shows that
they have been recently formed -- in other words, star formation is a
process which takes place even today.

These results revolutionised our understanding of stars and their evolution.
Presently they are accepted by the astronomical community.  Professor of
astronomy in Tartu University Rootsm\"ae (1885 - 1959) applied these ideas to
kinematics of stars to find the sequence of formation of different stellar
populations\cite{Rootsmae:1961}.  Similar ideas were developed independently
by Eggen, Lynden-Bell and Sandage\cite{Eggen:1962}.

\subsection{Grigori Kuzmin, the density of matter and galactic dynamics}  

\begin{figure}[ht]
\centering
\resizebox{.45\columnwidth}{!}{\includegraphics*{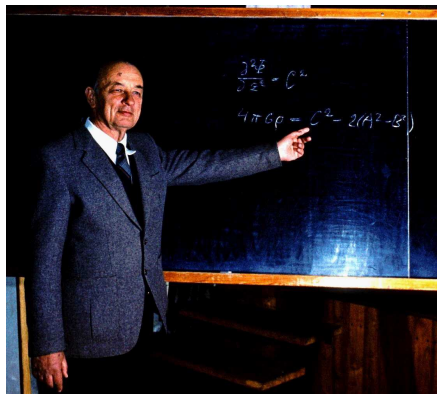}}
\caption{Grigori Kuzmin explaining his method to determine the density
of matter near the Sun}
\label{fig:kuzmin}
\end{figure}

\"Opik's student Grigori Kuzmin (1917 - 1988) continued his mentors work on
studying the structure of the Galaxy.  In his PhD thesis he developed further
\"Opik's method to derive the density of matter in the Galaxy and constructed
a mathematical model of the Galaxy, much more advanced than previous
ones\cite{Kuzmin:1952, Kuzmin:1952a, Kuzmin:1955, Kuzmin:1956}.  In contrast
to earlier models by Oort and Schmidt he used ellipsoids of variable density,
and applied the theory first to the Andromeda galaxy and then to our own
Galaxy.  One of central problems in modeling the Galaxy was the density of
matter near the Sun.  In 1930s famous Dutch astronomer Jan
Oort\cite{Oort:1932} studied this problem and found, that in addition to
ordinary stars there must be in the Galaxy an unknown population, so that the
total local density exceeds twice the density of visible matter.  Kuzmin found
that the method used by Oort is not very accurate and that there is no
indication for the presence of local dark matter in the Galaxy.  Later two
students of Kuzmin, Heino Eelsalu (1930 - 1998) \cite{Eelsalu:1959} and Mihkel
J\~oeveer (1937 - 2006) \cite{Joeveer:1972, Joeveer:1974}, reanalysed the
problem, using different data and methods, and confirmed his results.

The discrepancy between results of Tartu astronomers and the rest of the world
continued until 1990s, when finally modern data confirmed that Kuzmin was
right (Gilmore, Wyse \& Kuijken\cite{Gilmore:1989}). Thus we came to the
conclusion that {\em there is no evidence for the presence of large amounts of
  dark matter in the disk of the Galaxy}.  If there is some invisible matter
near the galactic plane, then it consists probably of low--mass stars or
jupiters, which have been formed from the flat gas population.

Kuzmin also developed a method how to describe more accurately the
kinematics of stellar populations, using three integrals: the energy
and mass conservation integrals, and a third integral, which allows the
existence of three-axial velocity ellipsoids of stellar
populations\cite{Kuzmin:1952a, Kuzmin:1956}.  From his model it
follows that orbits of stars in galaxies lie in a toroidal volume.
Numerical modeling of star orbits in realistic mass models have
confirmed Kuzmin's prediction.

\section{Following \"Opik and Kuzmin: detailed galactic models}

\subsection{Methods of galactic modeling}

The work by \"Opik, Rootsm\"ae, Kuzmin and their students formed the
basis of a concept of the structure and evolution of stellar
populations in galaxies, which is rather close to the presently
accepted picture.  In early 1960s I was interested in the problem too.
As new observational data arrived, the need for a better and more
accurate model of our Galaxy and other galaxies was evident.  Detailed
local structure is known only for our own Galaxy, and global
information on stellar populations is better known for external
galaxies, thus it is reasonable to investigate the structure of our
Galaxy and other galaxies in parallel.  

Also there was a need for a more detailed method of the construction of
composite models of galaxies.  This goal was realised in a series of papers in
Tartu Observatory Publications\cite{1968PTarO..36..341K, 1968PTarO..36..357E,
  1968PTarO..36..396E, 1968PTarO..36..414E, 1968PTarO..36..442E}, a summary of
the method was published in English\cite{Einasto:1969kt}.  A natural
generalisation of classical galactic models is the use of all available
observational data for spiral and elliptical galaxies, both photometric data
on the distribution of colour and light, and kinematical data on the rotation
and/or velocity dispersion.  Further, it is natural to apply identical methods
for modeling of galaxies of different morphological type (including our own
Galaxy), and to describe explicitly all major stellar populations.

The main principles of model construction were: (1) galaxies can be considered
as sums of physically homogeneous populations (young flat disk, thick disk,
core, bulge, halo); (2) physical properties of populations (mass--to--luminosity
ratio $M/L$, colour) should be in agreement with models of physical evolution of
stellar populations; (3) the density of a population can be expressed as
ellipsoids of constant flatness and rotational symmetry; (4) densities of
populations are non--negative and finite; (5) moments of densities which
define the total mass and effective radius of the galaxy are finite.  It was
found that in a good approximation densities of all stellar populations can be
expressed by a generalised exponential law: $\rho (a)=\rho (0)\exp
[-(a/a_c)^{1/N}]$, where $\rho (0)=hM/(4\pi\epsilon a_0^3)$ is the central
density, $a= \sqrt{R^2+z^2/\epsilon^2}$ is the distance along the major axis,
$\epsilon$ is the axial ratio of the equidensity ellipsoid, $a_c=ka_0$ is the
core radius ($a_0$ is the harmonic mean radius), $h$ and $k$ are normalising
parameters, depending on the structural parameter $N$, which allows to vary
the density behaviour with $a$. The cases $N=1$ and $N=4$ correspond to
conventional exponential and de Vaucouleurs models, respectively.

This density law (called Einasto profile) was first applied to find a
composite model of the Galaxy\cite{1965TarOT..17..1E}, based on the
new system of Galactic constants, using all available data (Einasto
and Kutuzov\cite{1964TarOT..11..11E}). Next the method was applied to
the Andromeda galaxy \cite{1969Ap......5...67E, 1970Ap......6...69E,
  1970Ap......6..120E}.

\begin{figure*}[ht]
\centering
\resizebox{.45\columnwidth}{!}{\includegraphics*{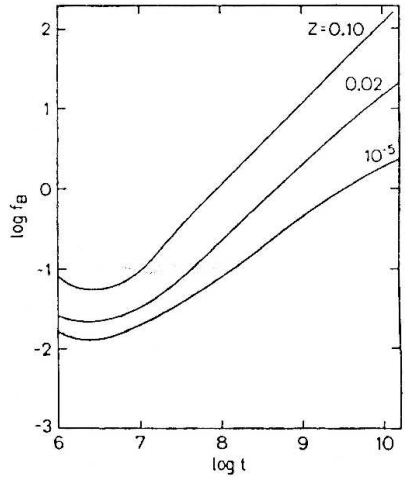}}
\caption{The evolution of mass-to-luminosity ratios $f_B=M/L$ for populations
  of various age $t$ and metal content $Z$, according to model
  calculations\cite{1974smws.proc..291E}.  }
\label{fig:evolution}
\end{figure*}

A central problem in galactic modeling is the correct estimation of the mass-to-luminosity ratio of populations.  This ratio depends on the evolutionary history of the population and on its chemical composition.  In order to bring these ratios for different populations to a coherent system, a model of physical evolution of stellar populations was developed \cite{1972thesis..E, 1974smws.proc..291E}.  The model used as input data the evolutionary tracks of stars of various composition (metallicity) and age; the star formation rate as a function of stellar mass was accepted according to Salpeter\cite{Salpeter:1955}, with a low--mass limit of star formation of $M_0 \approx 0.03~M_{\odot}$. The model yielded a continuous sequence of population parameters as a function of age (colour, spectral energy distribution, $M/L$), see Fig.~\ref{fig:evolution}.  The results of modeling stellar populations were calibrated using direct dynamical data for star clusters and central regions of galaxies (velocity dispersions) by Einasto \& Kaasik\cite{1973ATsir.790....1E}.  These data supported relatively high values ($M/L\approx 10 - 30$) for old metal--rich stellar populations near centres of galaxies; moderate values ($M/L\approx 3 - 10$) for discs and bulges; and low values ($M/L\approx 1 - 3$) for metal--poor halo--type populations.  Modern data yield lower values, due to more accurate measurements of velocity dispersions in clusters and central regions of galaxies, and rotation data on bulge dominated S0 galaxies.

\section{Dark matter in galaxies}

\subsection{Masses and radii of galaxies}

I had a  problem in the modeling of M31.  If rotation data were
taken at face value, then it was impossible to represent the
rotational velocity with the sum of known stellar populations. The local
value of $M/L$ increases towards the periphery of M31 very rapidly, if the
mass distribution is calculated directly from rotation velocity.  All
known old metal--poor halo--type stellar populations have a low $M/L
\approx 1 - 3$.  In contrast, on the basis of rotation data we got $M/L >
1000$ on the periphery of the galaxy near the last point with measured
rotational velocity.

I discussed the problem with my collaborator Enn Saar.  He suggested to abandon the idea, that only stellar populations exist in galaxies. Instead it is reasonable to assume the existence of a population of unknown nature and origin, and to look which properties it should have using available data on known stellar populations. So I calculated a new set of models for M31, our Galaxy and several other galaxies of the Local Group, as well as for the giant elliptical galaxy M87 in the Virgo cluster. In most models it was needed to include a new population in order to bring rotation and photometric data into mutual agreement. To avoid confusion with the metal-poor halo the new hypothetical population was called  corona.

\begin{figure*}[ht]
\centering
\resizebox{.48\columnwidth}{!}{\includegraphics*{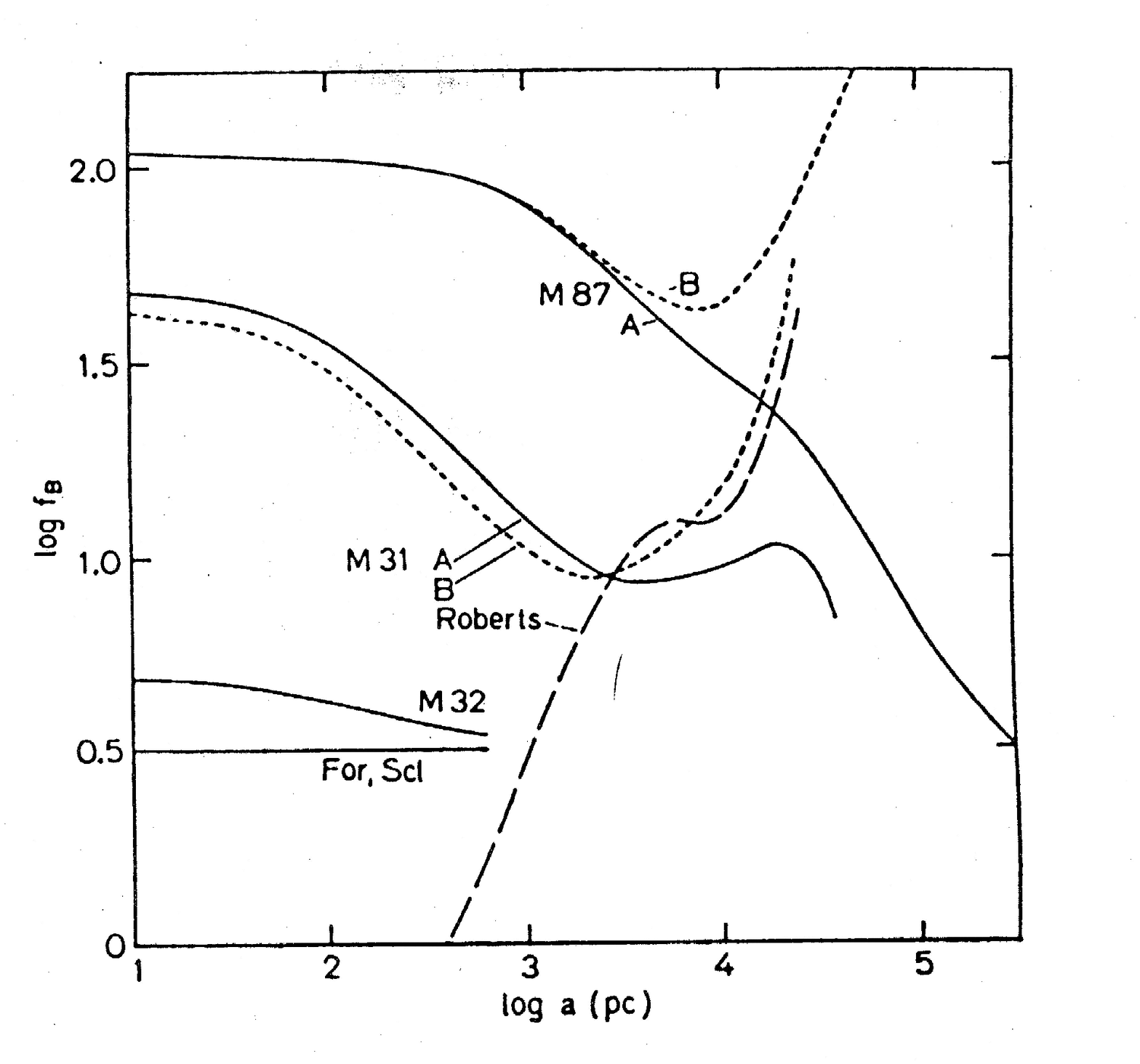}}
\resizebox{.46\columnwidth}{!}{\includegraphics*{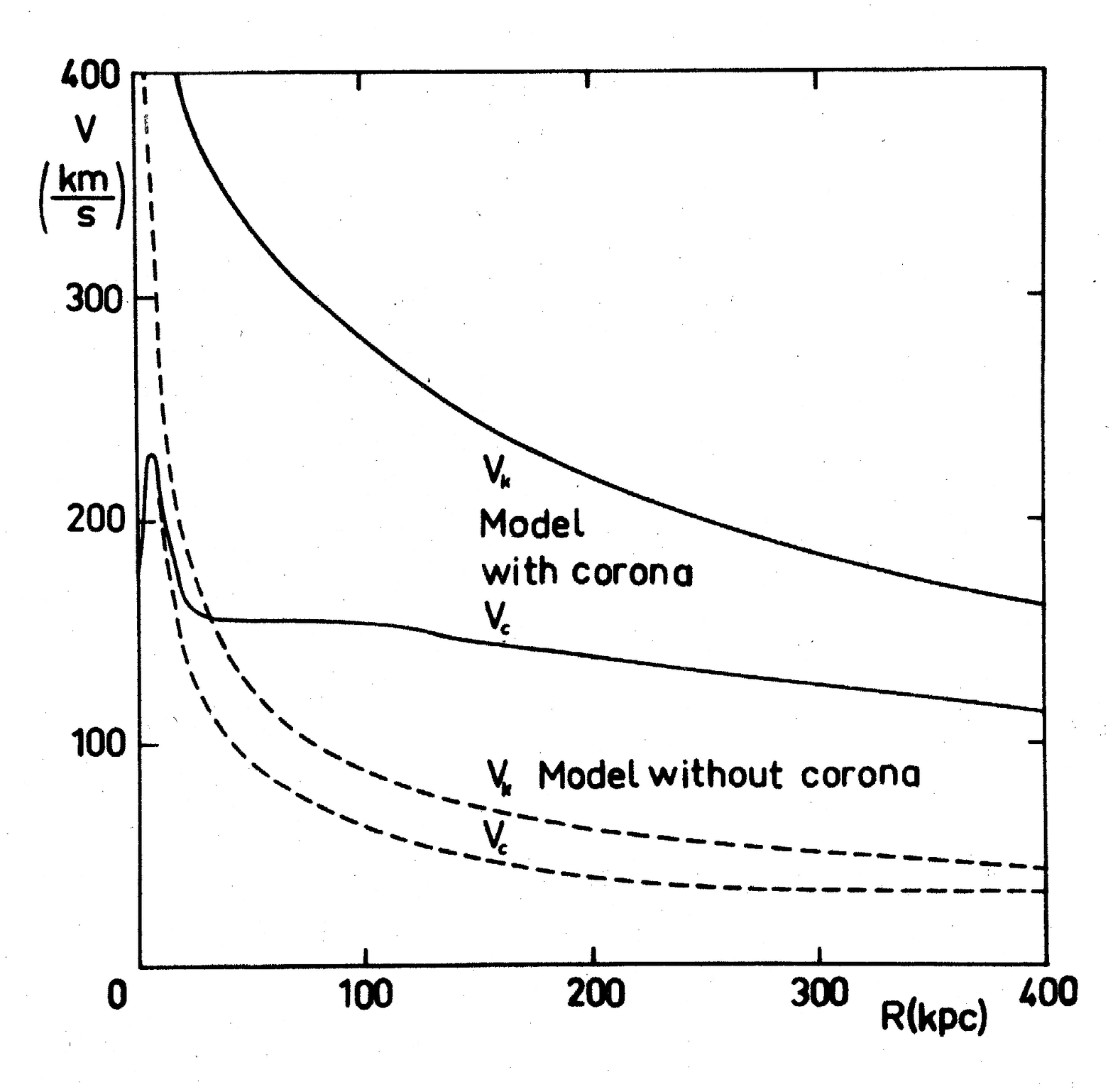}}
\caption{Combined physical and evolution models of M31, M32, M87, For,
  Scl. Left: mass-to-luminosity ratio as a function of radius. Models without
  and with massive corona are shown, cases A and B,
  respectively\cite{1974smws.proc..291E}. Right: circular and escape
  velocities of for a model of the Galaxy with and without a
  corona\cite{1976sgov.meet..398E}. }
\label{fig:M31model}
\end{figure*}

Results of these calculations were reported in the First European Astronomy
Meeting in Athens in September 1972\cite{1974smws.proc..291E}.  The
conclusions were: (1) There are two dark matter (DM) problems: the local DM
near the Galactic plane, and the global DM forming an extended almost
spherical population (corona); (2) The local dark matter, if it exists, must
be of stellar origin, as it is strongly concentrated to the Galactic plane;
(3) The global dark matter is probably of non-stellar origin. Available data
were insufficient to determine outer radii and masses of coronas. Preliminary
estimates indicated that in some galaxies the mass and radius of the corona
may exceed considerably the mass and radius of known stellar populations.

Arguments for the non-stellar origin of galactic coronas were the following. (1) Physical and kinematical properties of the stellar populations depend almost continuously on the age of the population, the oldest have the lowest metallicity and $M/L$-ratio, and there is no place where to put the corona into this sequence\cite{1974smws.proc..291E}. (2) Since the $M/L$ value and spatial distribution of the corona differ so much of similar properties of known stellar populations, the corona must have been formed much earlier than all known populations; the total mass of the corona exceeds masses of known populations by an order of magnitude, thus we have a problem: How to transform in an early stage of the evolution of the universe most of gas to the coronal stars? It is known that star formation is a very inefficient process, as in a star-forming gaseous nebula only about 1~\% of matter transforms to stars\cite{1974smws.proc..291E}. (3) Due to the large size of the corona, coronal stars must have in the vicinity of the Sun much higher velocities than all other stars, but no extremely high-velocity stars have been found by Jaaniste and Saar\cite{Jaaniste:1975}. (4) Luminosity decreases in outer regions rapidly, therefore, if the matter is in stars, they must be of very low luminosity. The presence of low-luminosity stars in outer galactic regions without bright ones would require a process of large-scale segregation of stars according to mass (low-luminosity stars have small masses), but this is highly improbable\cite{1974smws.proc..291E}.  The hidden matter cannot be in the form of neutral gas, since this gas would be observable\cite{1974smws.proc..291E}. For these reasons, I assumed that coronas may consist of hot  gas. Soon it was clear that a fraction of coronal matter is indeed gaseous\cite{1974Natur.252..111E}, however not all.

To find the radii and masses of galactic coronas more distant test objects are
needed.  One possibility is the use of companion galaxies. If coronas are
large enough, then in pairs of galaxies the companion galaxy is moving inside
the corona, and it can be considered as a test particle to measure the
gravitational attraction of the main galaxy.  Mean relative velocities,
calculated for different distances from the main galaxy, can be used instead
of rotation velocities to find the mass distribution of giant galaxies. A
collection of 105 pairs of galaxies yield following results: radii and masses
of galactic coronas exceed radii and masses of stellar populations of
galaxies by an order of magnitude!  Together with A. Kaasik and E. Saar we
calculated new models of galaxies including dark coronas. Results were
reported in the Caucasus Winter School in January 1974, and published in
Nature\cite{1974Natur.250..309E}. Our data suggest that all giant galaxies
have massive coronas of some unknown origin (Dark Matter), the total masses of
galaxies including dark coronas exceed masses of known populations by an
order of magnitude. It follows, that dark matter is the dominating component in
the whole universe.  Similar results have been obtained by Ostriker, Peebles
and Yahil\cite{Ostriker:1974}. Additional arguments supporting the physical
connection between main galaxies and their companions were found from the
morphology of companions\cite{1974Natur.252..111E}.

\begin{figure*}[ht]
\centering
\resizebox{.50\columnwidth}{!}{\includegraphics*{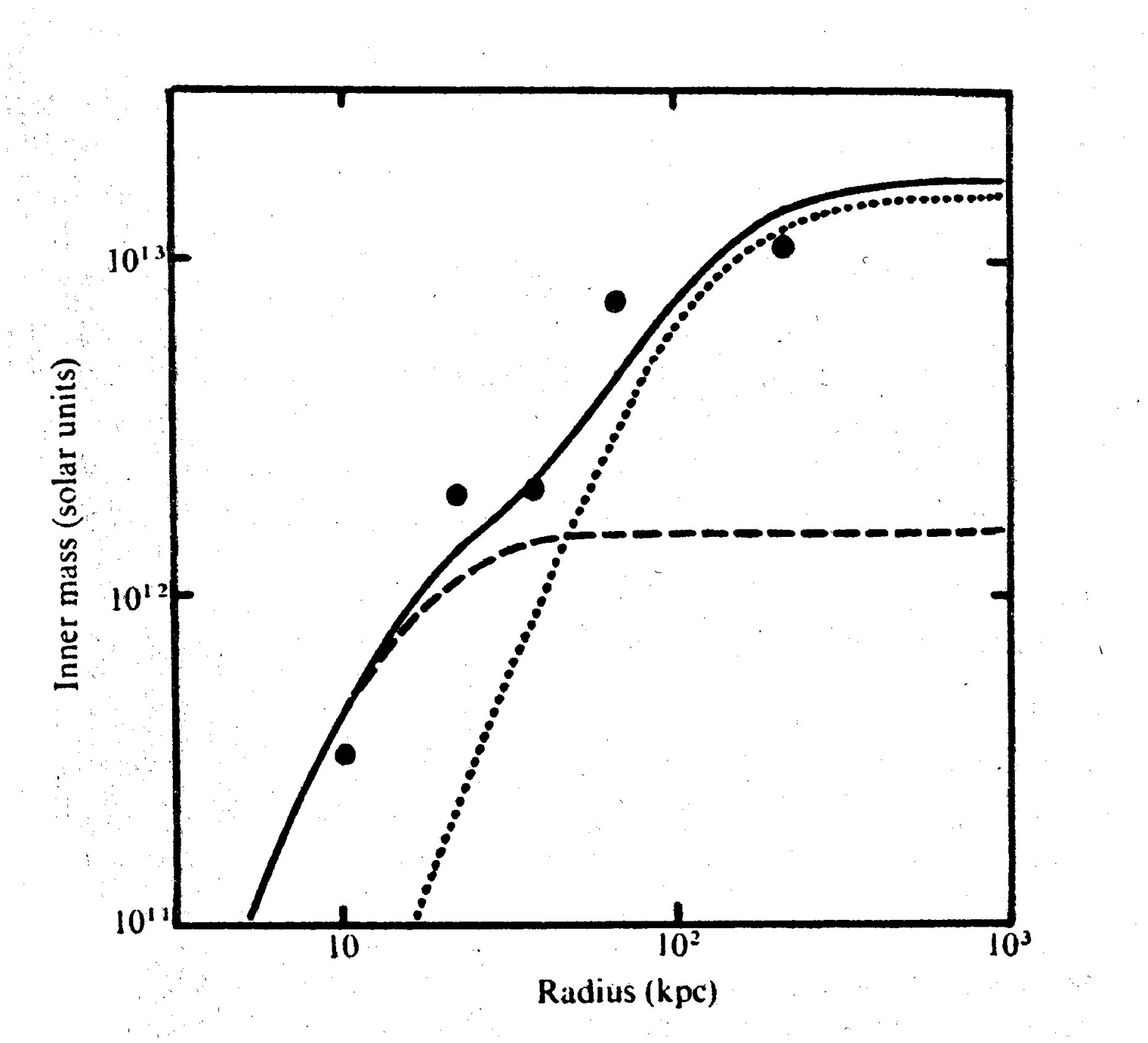}}
\caption{Internal mass of galaxies: dots - data on 105 companion
  galaxies; dashed line - mass due to known galactic populations;
  dotted line - mass due to dark corona; solid line - total internal
  mass from models\cite{1974Natur.250..309E}. }
\label{fig:intmass}
\end{figure*}

\subsection{Problems with dark matter}

In January 1975 the first conference on dark matter was held in Tallinn,
Estonia.  The rumour on dark matter had spread around the astronomical
community and, in contrast to conventional local astronomy conferences,
leading Soviet astronomers and physicists attended.  The main topics was the
possible nature of the dark matter.  It was evident that a stellar origin is
almost excluded\cite{Jaaniste:1975}, but a fully gaseous corona also has
difficulties, as shown by Komberg and Novikov\cite{Komberg:1975}. The problems
with baryon nucleosynthesis constraints were discussed by Zeldovich. So the
nature of DM was not clear.

The next dark matter discussion was in July 1975 during the Third
European Astronomical Meeting in Tbilisi, Georgia, where a full
session was devoted to the dark matter problem. Here the principal
discussion was between the supporters of the classical paradigm with
conventional mass estimates of galaxies, and of the new one with dark
matter.  The major arguments supporting the classical paradigm were
summarised by Gustav Tammann\cite{Materne:1976}. The most serious
arguments were: {\em Big Bang nucleosynthesis suggests a low-density
  Universe with the density parameter $\Omega \approx 0.05$; the
  smoothness of the Hubble flow also favours a low-density Universe.}

Dark matter problem was also discussed during the IAU General
Assembly in Grenoble, 1976.  Here arguments for the non--stellar
nature of dark coronas were again presented\cite{1976TarOT..54....3E}. 

\begin{figure*}[ht]
\centering
\resizebox{.49\columnwidth}{!}{\includegraphics*{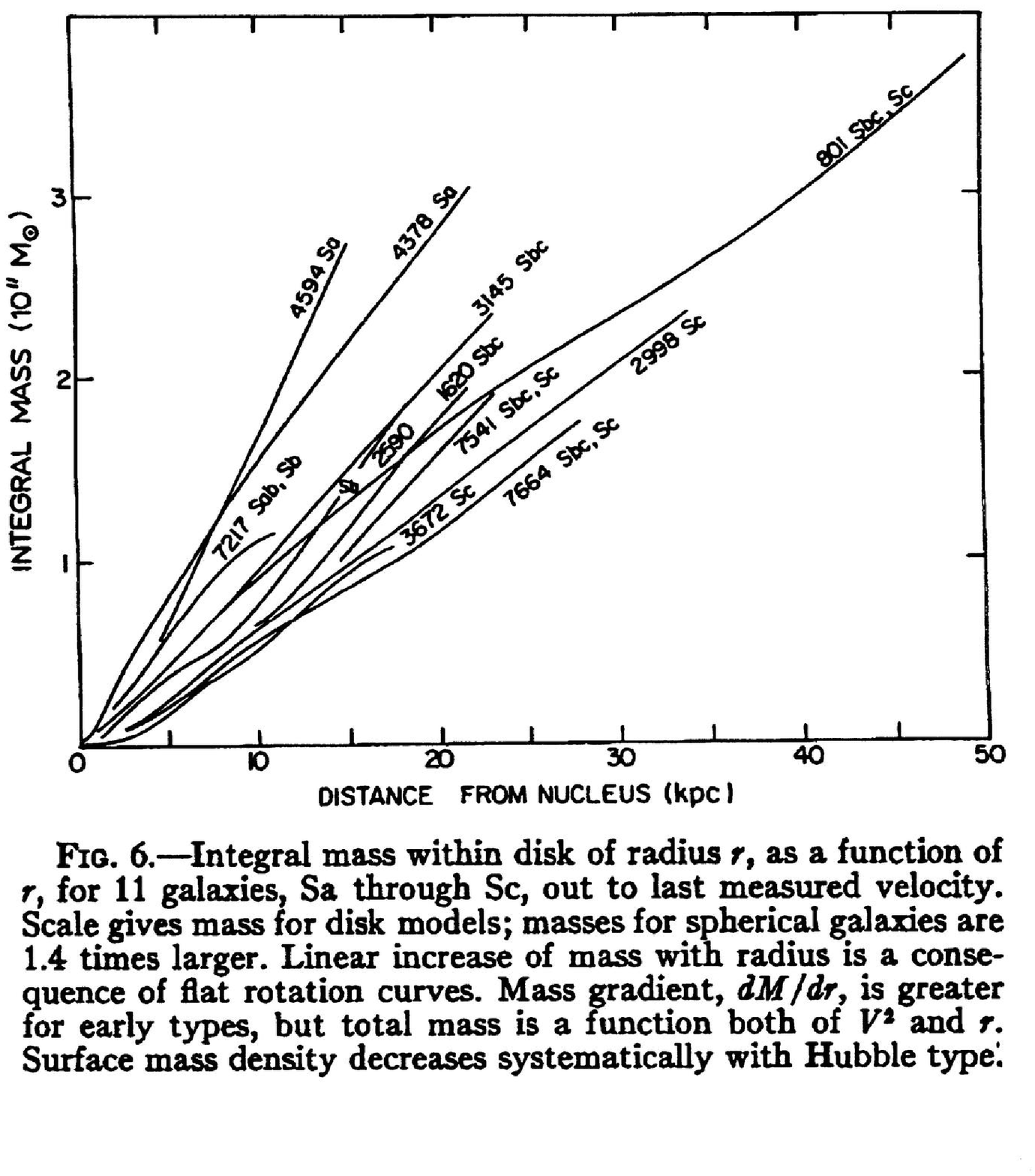}}
\resizebox{.46\columnwidth}{!}{\includegraphics*{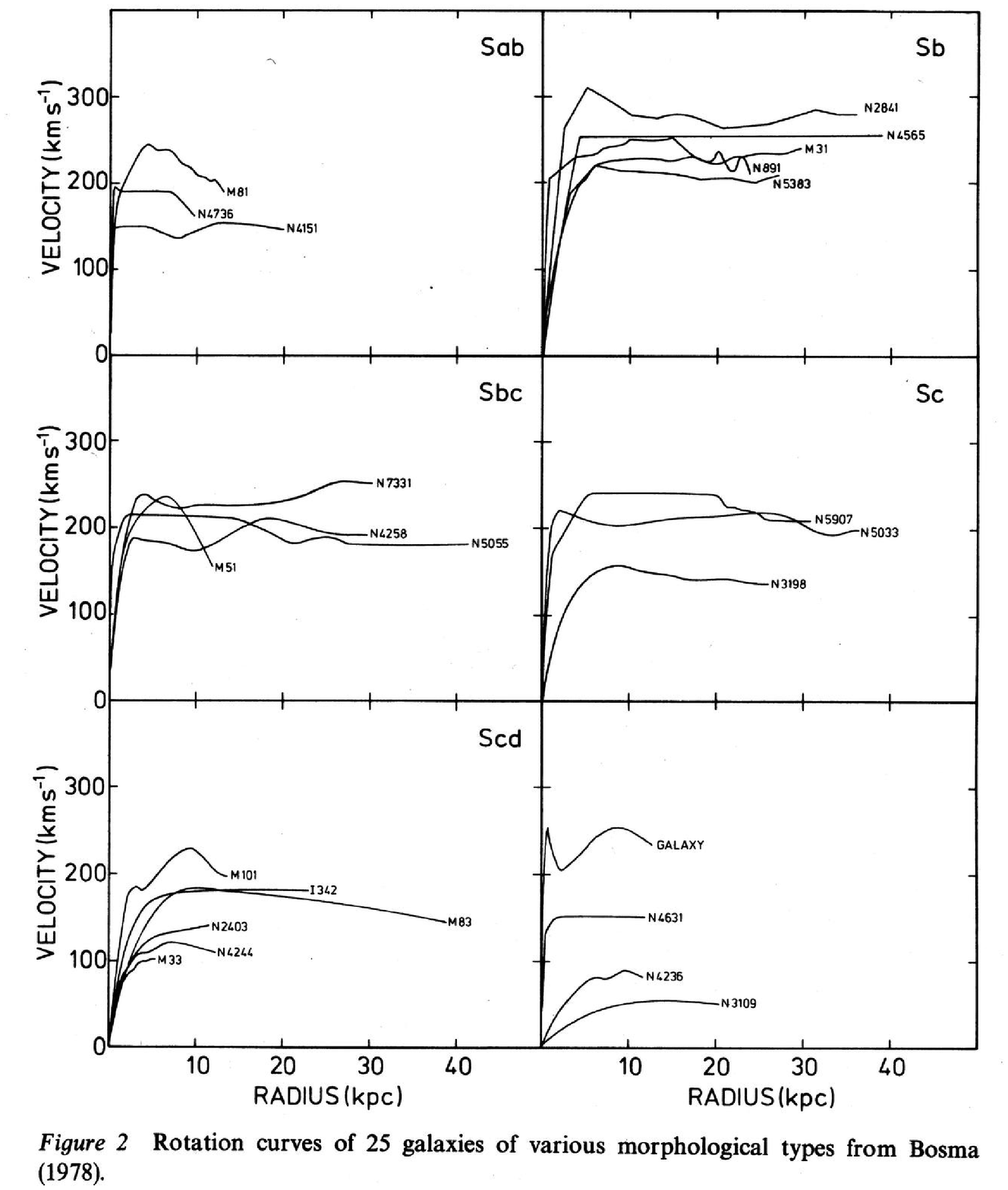}}
\caption{Left: Integral masses of spiral galaxies, derived by Rubin et
  al.\cite{Rubin:1978}.  Right: rotation curves of 25 spiral galaxies by
  Bosma\cite{Bosma:1978}. }
\label{fig:bosma_rubin}
\end{figure*}

It was clear that by sole discussion the presence and nature of dark
matter cannot be solved, new data and more detailed studies were
needed. A very strong confirmation of the dark matter hypothesis came
from new extended rotation curves of galaxies.  Vera Rubin and her
collaborators developed new sensitive detectors to measure optically
the rotation curves of galaxies at very large galactocentric
distances.  Their results suggested that practically all spiral
galaxies have extended flat rotation curves\cite{Rubin:1978,
  Rubin:1980}.  The internal mass of galaxies rises with distance
almost linearly, up to the last measured point, see
Fig.~\ref{fig:bosma_rubin}.  At the same time measurements of a number
of spiral galaxies with the Westerbork Synthesis Radio Telescope were
completed, and mass distribution models were built, all-together for
25 spiral galaxies by Bosma\cite{Bosma:1978}, see Fig.~\ref{fig:bosma_rubin}.
Observations confirmed the general trend that the mean rotation curves
remain flat over the whole observed range of distances from the
center, up to $\sim 40$ kpc for several galaxies. 

These new observations confirmed the presence of dark coronas of
galaxies. However, the nature of the coronas was still unclear, and
the difficulties discussed in Tallinn and Tbilisi were not clarified.

\subsection{Non-baryonic nature of Dark Matter}

In late 1970s suggestions were made that some sort of non-baryonic
elementary particles may serve as candidates for dark matter
particles.  There were several reasons to search for non-baryonic
particles as a dark matter candidate.  First of all, no baryonic
matter candidate did fit the observational data.  Second, the total
amount of dark matter is of the order of 0.2--0.3 in units of the
critical cosmological density, whereas the nucleosynthesis constraints
suggest that the amount of baryonic matter cannot be higher than about
0.04 of the critical density.

A very important observation was made, which caused doubts to the baryonic matter as the dark matter candidate. In 1964 Cosmic Microwave Background (CMB) radiation was detected.  Initially the Universe was very hot and all density and temperature fluctuations of the primordial gas were damped by very intense radiation. At a certain epoch called recombination the gas became neutral, and density fluctuations in the gas had a chance to grow by gravitational instability.  But gravitational clustering is a very slow process.  In order to have time to build up all observed structures the amplitude of initial density fluctuations at the epoch of recombination must be of the order of $10^{-3}$ of the density itself.  Density fluctuations are of the same order as temperature fluctuations, and astronomers started to search for temperature fluctuations of the CMB radiation. None were found. As the accuracy of measurement increased, lower and lower upper limits for the amplitude of CMB fluctuations were obtained.  In late 1970s it was clear that the upper limits are much lower than the theoretically predicted limit $10^{-3}$.

Then astronomers recalled the possible existence of non-baryonic particles, such as heavy neutrinos. This suggestion was made independently by several astronomers (\cite{Szalay:1976, Tremaine:1979, Doroshkevich:1980b, Chernin:1981}).  If dark matter is non-baryonic, then this helps to explain the paradox of small temperature fluctuations of cosmic microwave background radiation.  Density perturbations of non-baryonic dark matter start growing already during the radiation-dominated era, whereas the growth of baryonic matter is damped by radiation.  If non-baryonic dark matter dominates dynamically, the total density perturbations can have an amplitude of the order $10^{-3}$ at the recombination epoch, which is needed for the formation of the observed structure of the Universe.  The evolution of perturbations in a neutrino-dominated dark matter medium was discussed in a conference in Tallinn in April 1981 (this conference was probably the first one devoted to the astro--particle physics).

Numerical simulations made for a neutrino-dominated universe were made by a number of astronomers.  These calculations demonstrated some weak points in the scenario: large-scale structures (superclusters) form too late and have no fine structure as observed in the real Universe. A new scenario was suggested, among others, by Bond, Szalay \& Turner\cite{Bond:1982}; here hypothetical particles like axions, gravitinos or photinos play the role of dark matter.  Numerical simulations of structure evolution for neutrino and axion--dominated universe were made and analysed by Melott \etal \cite{Melott:1983}.  All quantitative characteristics (connectivity of the structure, multiplicity of galaxy systems, correlation function) of this new model fit the observational data well.  This model was called subsequently the Cold Dark Matter (CDM) model, in contrast to the neutrino--based Hot Dark Matter model. Presently the CDM model with some modifications is the most accepted model of the structure evolution (Blumenthal \etal \cite{Blumenthal:1984}).

\section{Large scale structure of the Universe}

\subsection{Zeldovich question}

After my talk in the Caucasus Winter School Zeldovich turned to me and
offered collaboration in the study of the universe.  He was developing
a theory of the formation of galaxies -- the pancake
theory\cite{Zeldovich:1970}; an alternative whirl theory was
suggested by Ozernoy, and a third theory of hierarchical clustering by
Peebles.  Zeldovich asked for our help in solving the question: Can we
find some observational evidence which can be used to discriminate
between these theories?

Initially we had no idea how we can help Zeldovich.  But soon we remembered
our previous experience in the study of galactic populations: kinematical and
structural properties of populations hold the memory of their previous
evolution and formation (Rootsm\"ae\cite{Rootsmae:1961}, Eggen, Lynden-Bell
\& Sandage\cite{Eggen:1962}). Random velocities of galaxies are of the order
of several hundred km/s, thus during the whole lifetime of the Universe
galaxies have moved from their place of origin only about 1~\Mpc\ (we use the
Hubble constant in units $H_0 = 100~h$ km~s$^{-1}$~Mpc$^{-1}$).  In other
words -- if there exist some regularities in the distribution of galaxies,
these regularities must reflect the conditions in the Universe during the
formation of galaxies.

\begin{figure*}[ht]
\centering
\resizebox{.47\columnwidth}{!}{\includegraphics*{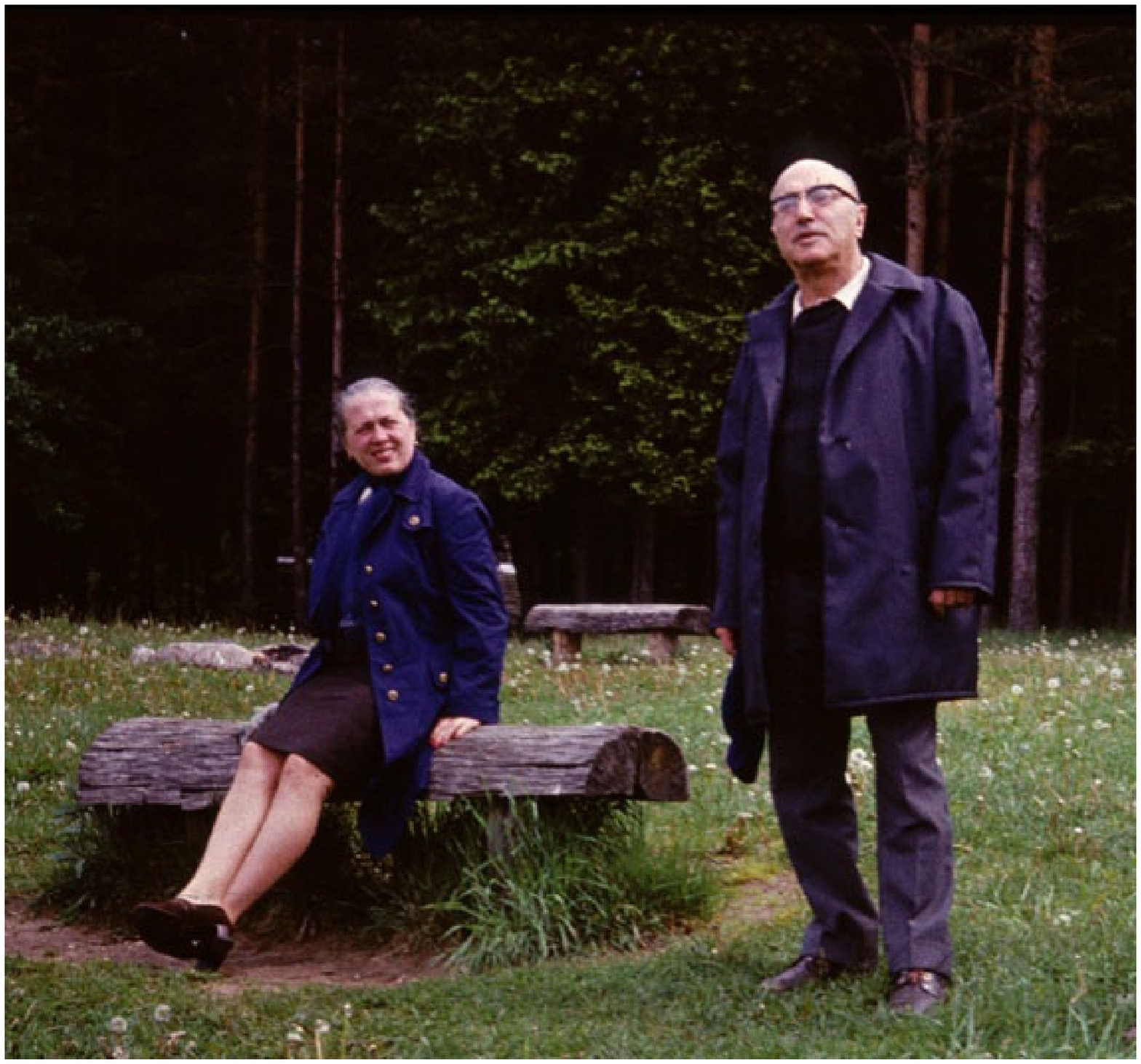}}
\resizebox{.45\columnwidth}{!}{\includegraphics*{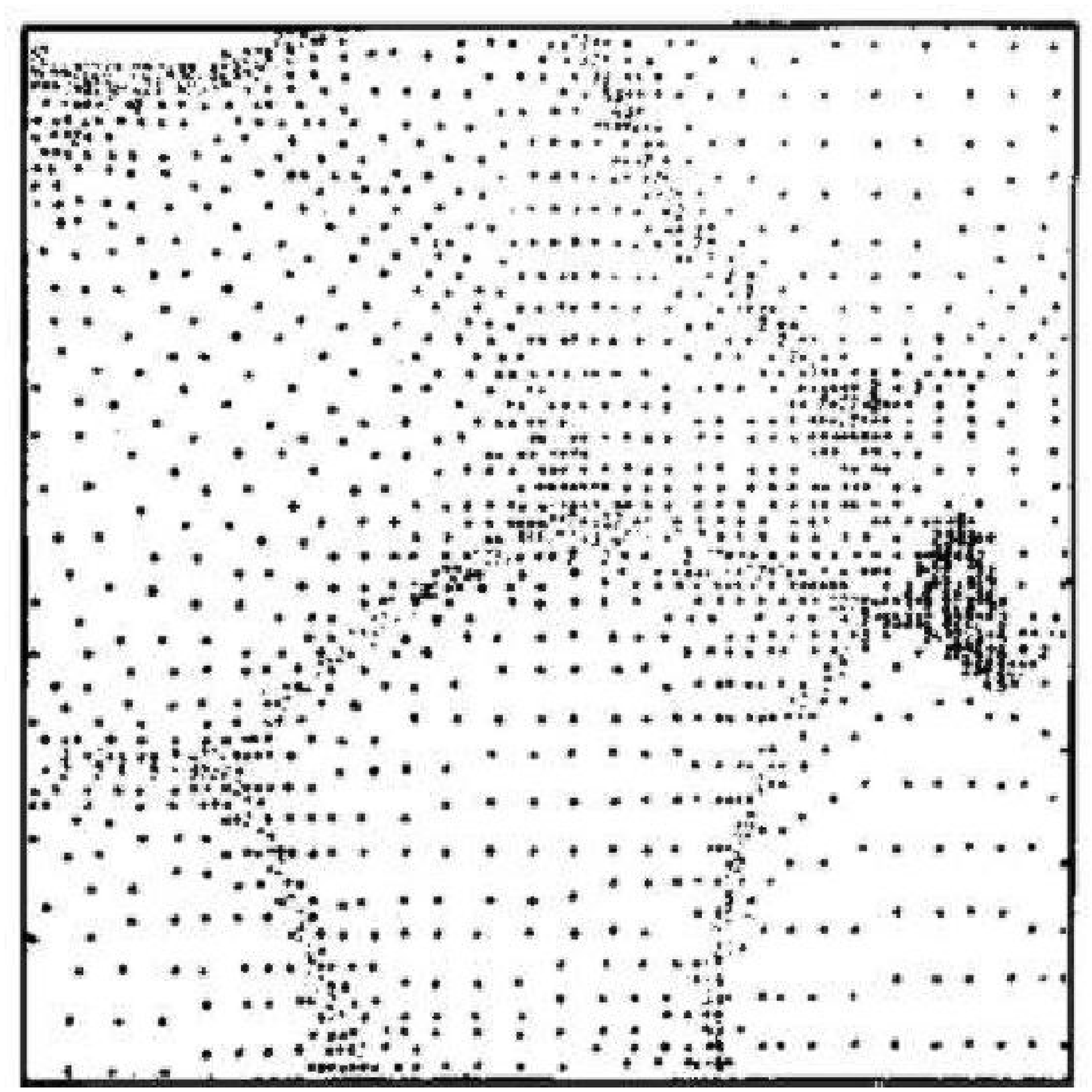}}
\caption{Left: Zeldovich with his wife visiting Estonia.
  Right: Distribution of particles in simulation (Zeldovich group 1975). }
\label{fig:zeld-nbody}
\end{figure*}

In our work to solve the Zeldovich question we had a close collaboration
with his team.  In 1975 Doroshkevich, Shandarin and
Novikov\cite{Doroshkevich:1980} obtained first results of numerical
simulations of the evolution of particles according to the theory of
gravitational clustering, developed by Zeldovich\cite{Zeldovich:1970}.
This was a 2--dimensional simulation with $128\times 128$ particles
(see Fig.~\ref{fig:zeld-nbody}). In this picture a system of high--
and low--density regions was seen: high--density regions form a
cellular network, which surrounds large under--dense regions.  One of
our challenges was to find out, whether the real distribution of
galaxies showed some similarity with the theoretical picture.

Now we had a guiding idea how to solve the problem of galaxy formation: {\em
  We have to study the distribution of galaxies on larger scales}.

\subsection{Tallinn Symposium on Large Scale Structure of the
  Universe}

Both our galactic astronomy and theoretical cosmology teams participated in
the effort to find the distribution of galaxies and their systems in space.
One approach was the study of the distribution of nearby Zwicky clusters.
Many bright galaxies of nearby Zwicky clusters had at this time measured
redshifts, so we hoped to determine the distribution of clusters and to find
some regularities there.  To see the distribution better, we built in the
office of Saar and Jaaniste a 3--dimensional model from plastic balls.  Some
regularity was evident: there were several clusters of Zwicky clusters --
superclusters, one of them in the Perseus region.  But too many clusters had
no galaxies with measured redshifts, so it was difficult to get an overall
picture. Zwicky nearby clusters were used several years later by Einasto,
J\~oeveer \& Saar\cite{Einasto:1980}, when more redshifts were available and
for the rest  photometric distances were estimated.

\begin{figure*}[ht]
\centering
\resizebox{.47\columnwidth}{!}{\includegraphics*{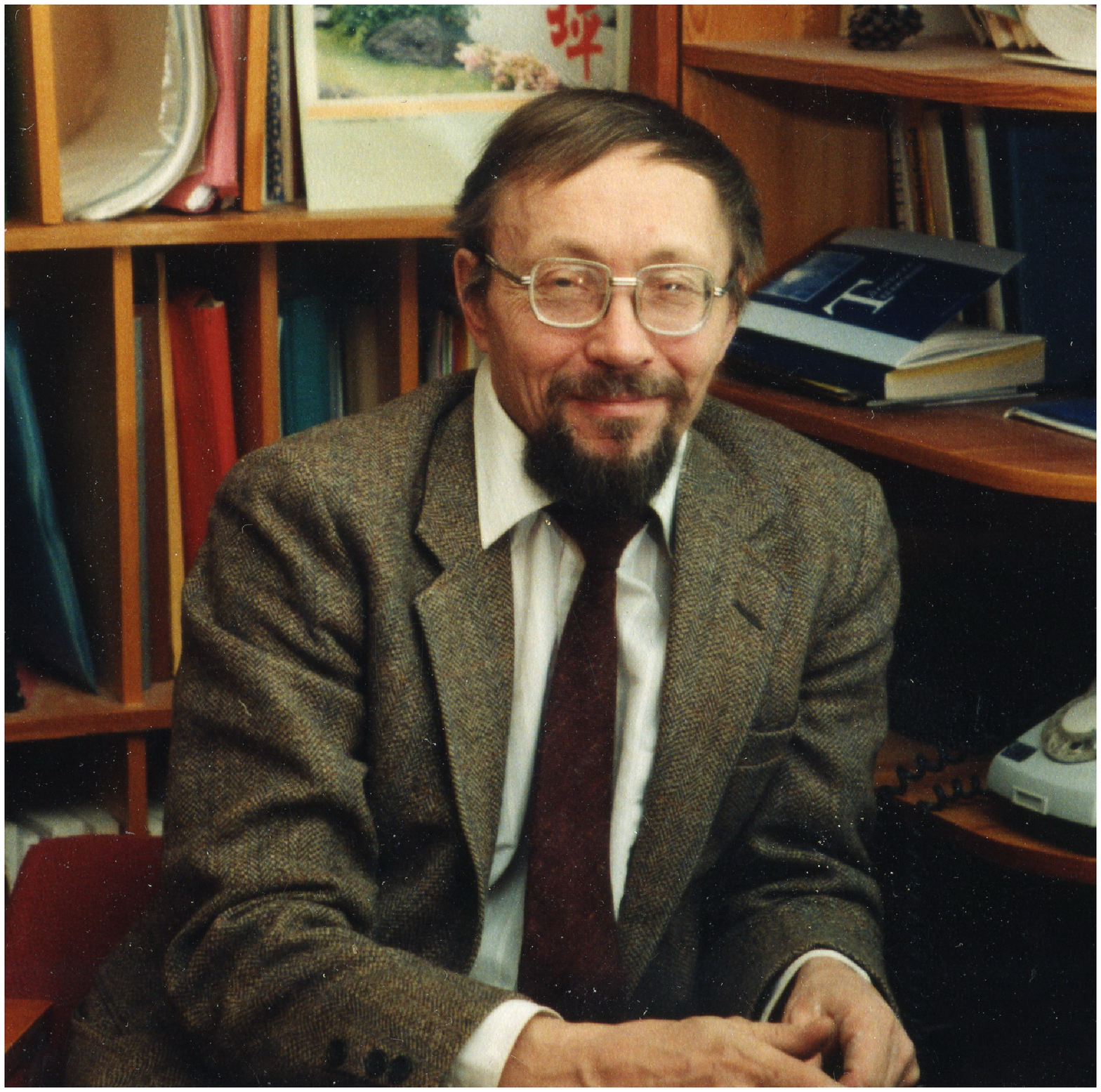}}
\resizebox{.44\columnwidth}{!}{\includegraphics*{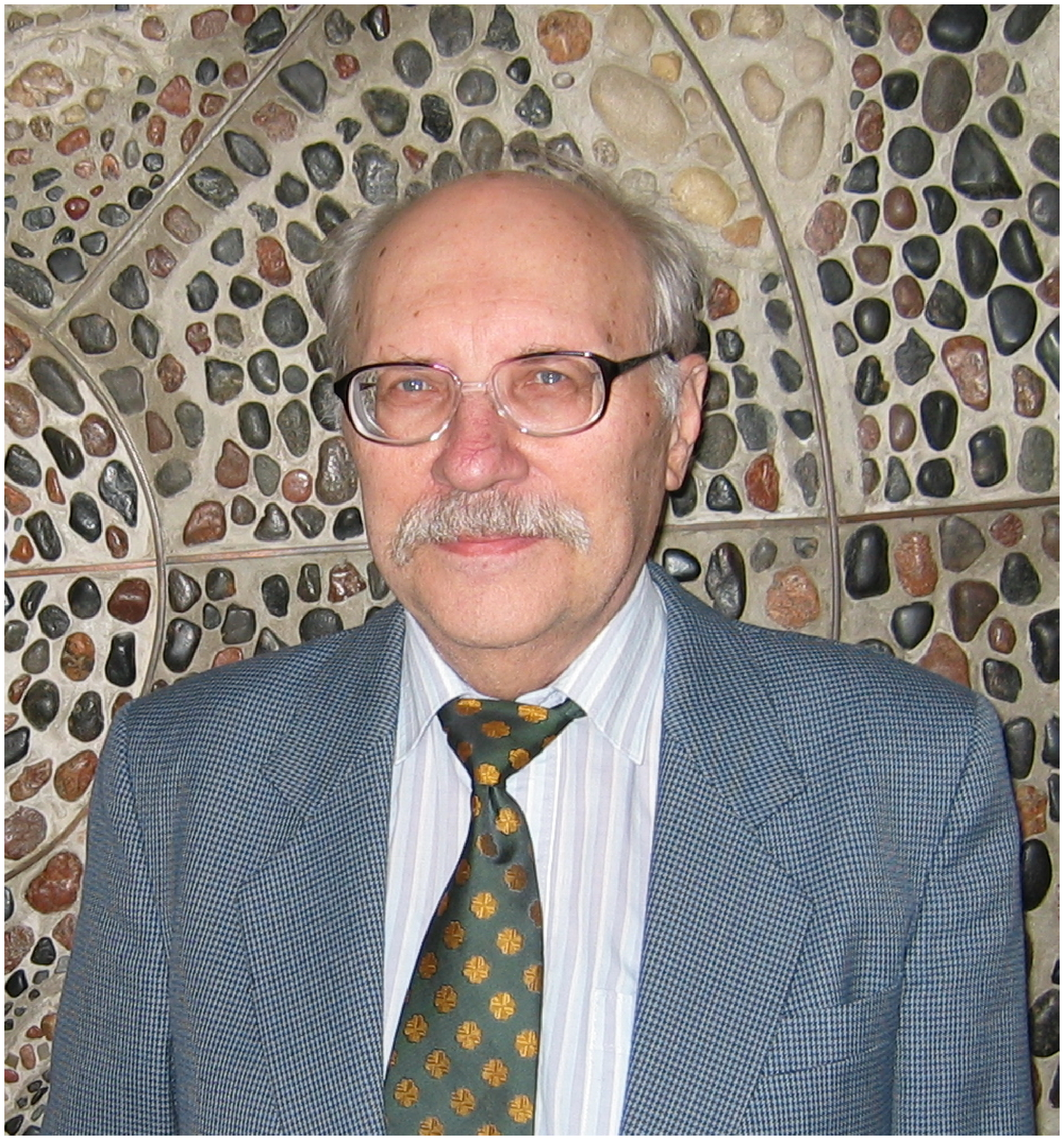}}
\caption{Left: Enn Saar. Right: Mihkel J\~oeveer.  }
\label{fig:saar}
\end{figure*}

\begin{figure*}[ht]
\centering
\resizebox{.45\columnwidth}{!}{\includegraphics*{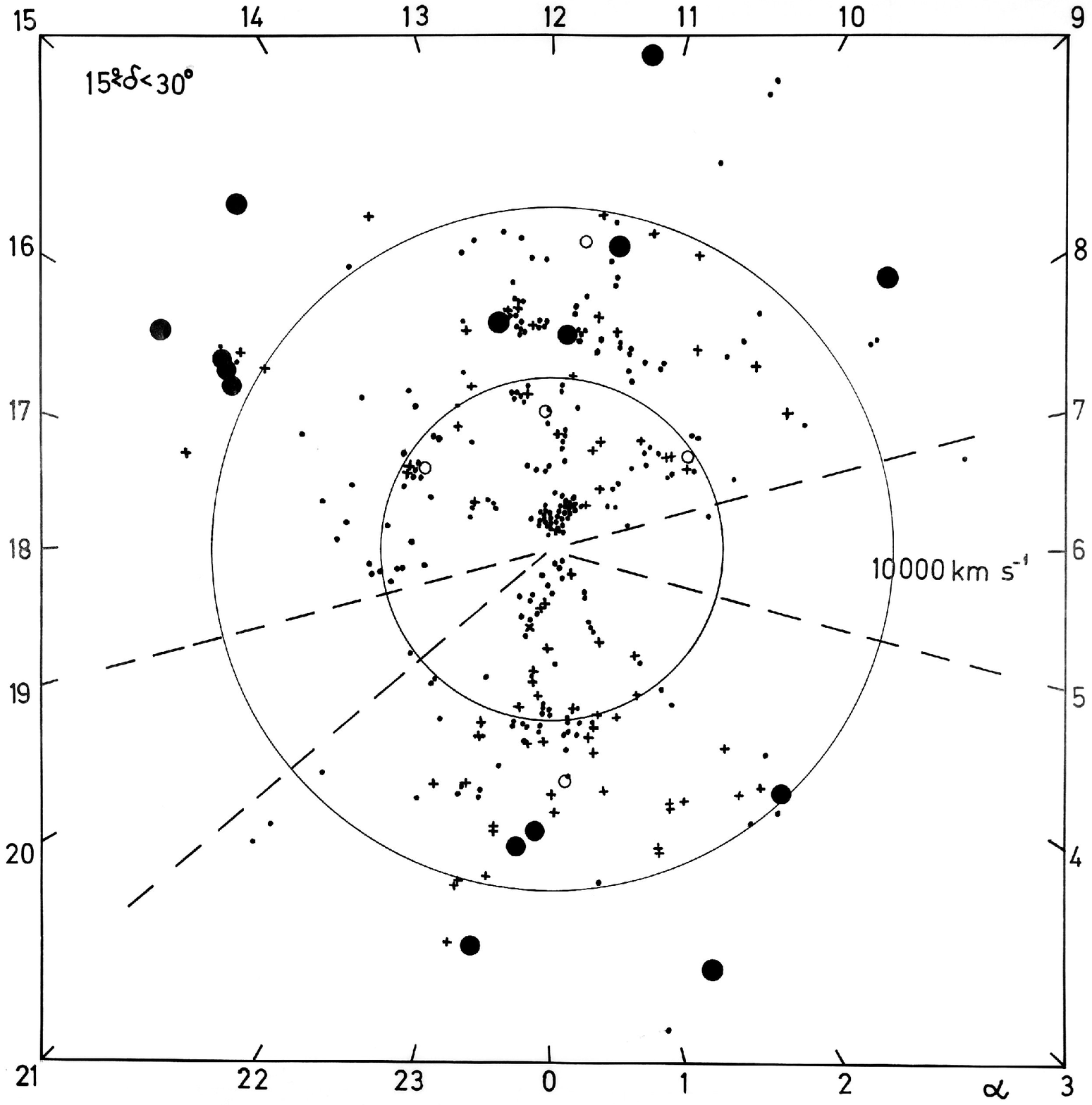}}
\resizebox{.45\columnwidth}{!}{\includegraphics*{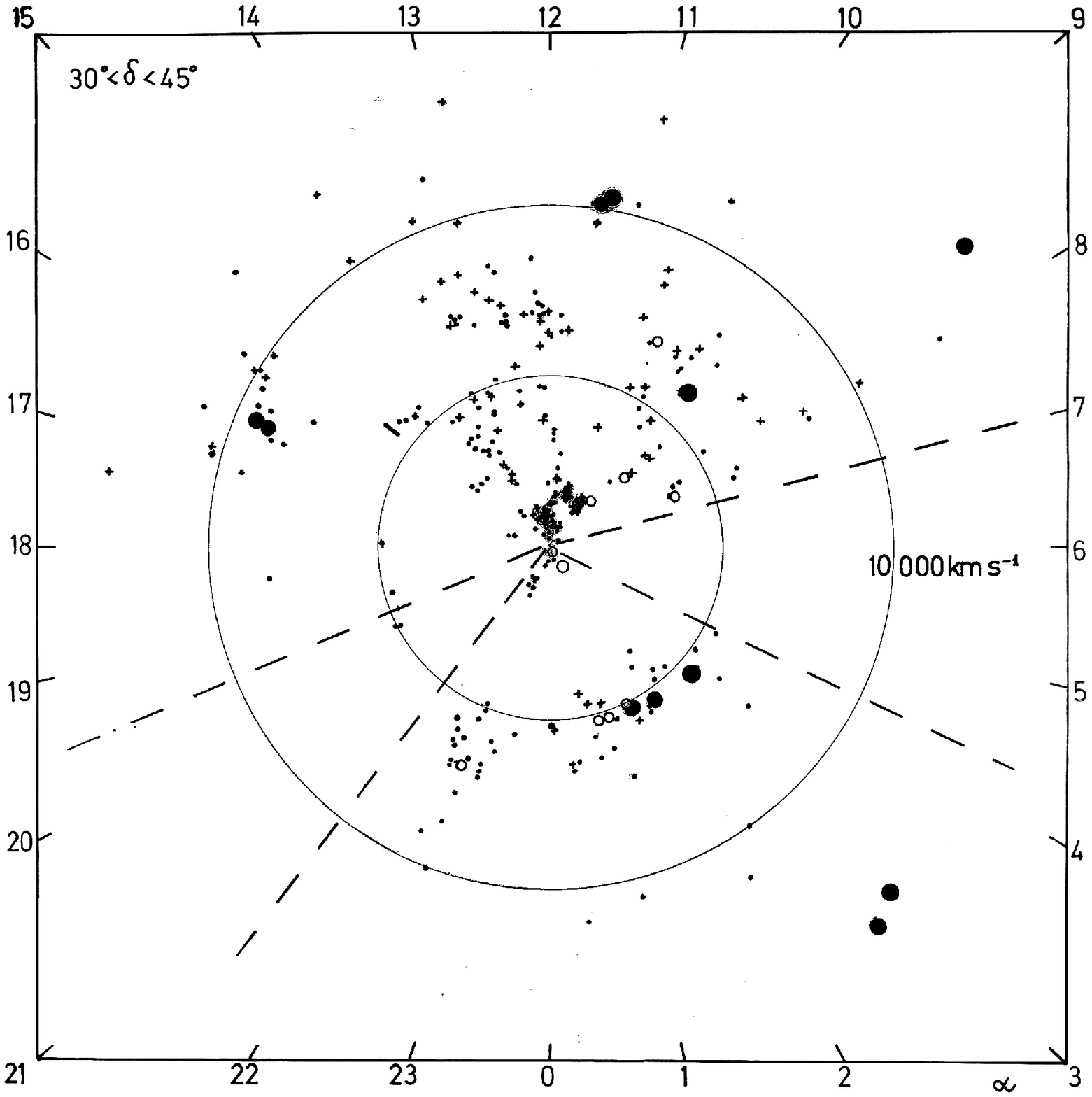}}
\caption{Clusters, groups and bright galaxies are mostly located in chains,
  concentrated to superclusters, weaker galaxy chains bind superclusters to
  infinite network. The space between galaxy-cluster chains is empty –- these
  regions are called holes or cosmic voids (J\~oeveer \&
  Einasto\cite{Joeveer:1978a})}.
\label{fig:pers}
\end{figure*}

A different approach was used by Mihkel J\~oeveer.  He used wedge--diagrams. His trick was: he made a number of wedge diagrams in sequence for fixed $\alpha$ and $\delta$ intervals, and plotted in the same diagram galaxies, as well as groups and clusters of galaxies, and Markarian galaxies.  Redshift data for clusters, groups and Markarian galaxies were almost complete in the Northern hemisphere up to a redshift about 15.000 km/s. Two wedge--diagrams of width $\Delta\delta =15^{\circ}$, crossing the Coma, Perseus, Hercules and Local superclusters, are shown in Fig.~\ref{fig:pers}.  In these diagrams a regularity was clearly seen: {\em
  isolated galaxies and galaxy systems populate identical regions, and the
  space between these regions is empty}. After this success the whole Tartu
cosmology team continued the study using wedge--diagrams and other methods to
analyse the distribution of galaxies and their systems. A detailed analysis of
the Perseus supercluster region was made; here the number of foreground
galaxies is very small.

Already in 1975, after the Tbilisi Meeting, we discussed with
Zeldovich the possibility to organise a real international conference
devoted solely to cosmology.  Due to Soviet bureaucratic system it was
extremely difficult for Soviet astronomers to attend international
conferences in Western countries; thus the only possibility to have a
better contact between cosmologists from East and West was to hold the
conference within the Soviet Union.  Zeldovich suggested to hold it in
Tallinn.  After some discussion we decided to devote it to ``Large
Scale Structure of the Universe''.  When we started preparations we
had no idea what this term could mean.

The symposium was held in September 1977.  Our main results were presented in
the talk by J\~oeveer \& Einasto\cite{Joeveer:1978a}: (1) galaxies, groups and
clusters of galaxies are not randomly distributed, but form chains,
concentrated in superclusters; (2) the space between galaxy chains contains
almost no galaxies and form holes or voids of diameter up to $\approx
70$~\Mpc\ (see Fig.~\ref{fig:pers}); (3) the whole pattern of the distribution
of galaxies and clusters resembles cells of a honeycomb, rather close to the
picture predicted by Zeldovich. A more detailed analysis was published
separately by J\~oeveer, Einasto \& Tago\cite{Joeveer:1978b}.

The presence of voids in the distribution of galaxies was reported also by
other groups: by Tully \& Fisher\cite{Tully:1978}, Tifft \&
Gregory\cite{Tifft:1978}, and Tarenghi \etal\cite{Tarenghi:1978} in the Local,
Coma and Hercules superclusters, respectively. Theoretical interpretation of
the observed cellular structure was discussed by
Zeldovich\cite{Zeldovich:1978}. Malcolm Longair noted in his concluding
remarks: {\em the discovery of the filamentary character of the distribution
  of galaxies, similar to a lace--tablecloth, and the overall cellular picture
  of the large--scale distribution was the most exciting result presented at
  this symposium}.  These results demonstrated that the pancake scenario by
Zeldovich\cite{Zeldovich:1970} has many advantages over other rivalling
scenarios.  The term ``Large--Scale Structure of the Universe'' got its
present meaning.

\begin{figure*}[ht]
\centering
\resizebox{.58\columnwidth}{!}{\includegraphics*{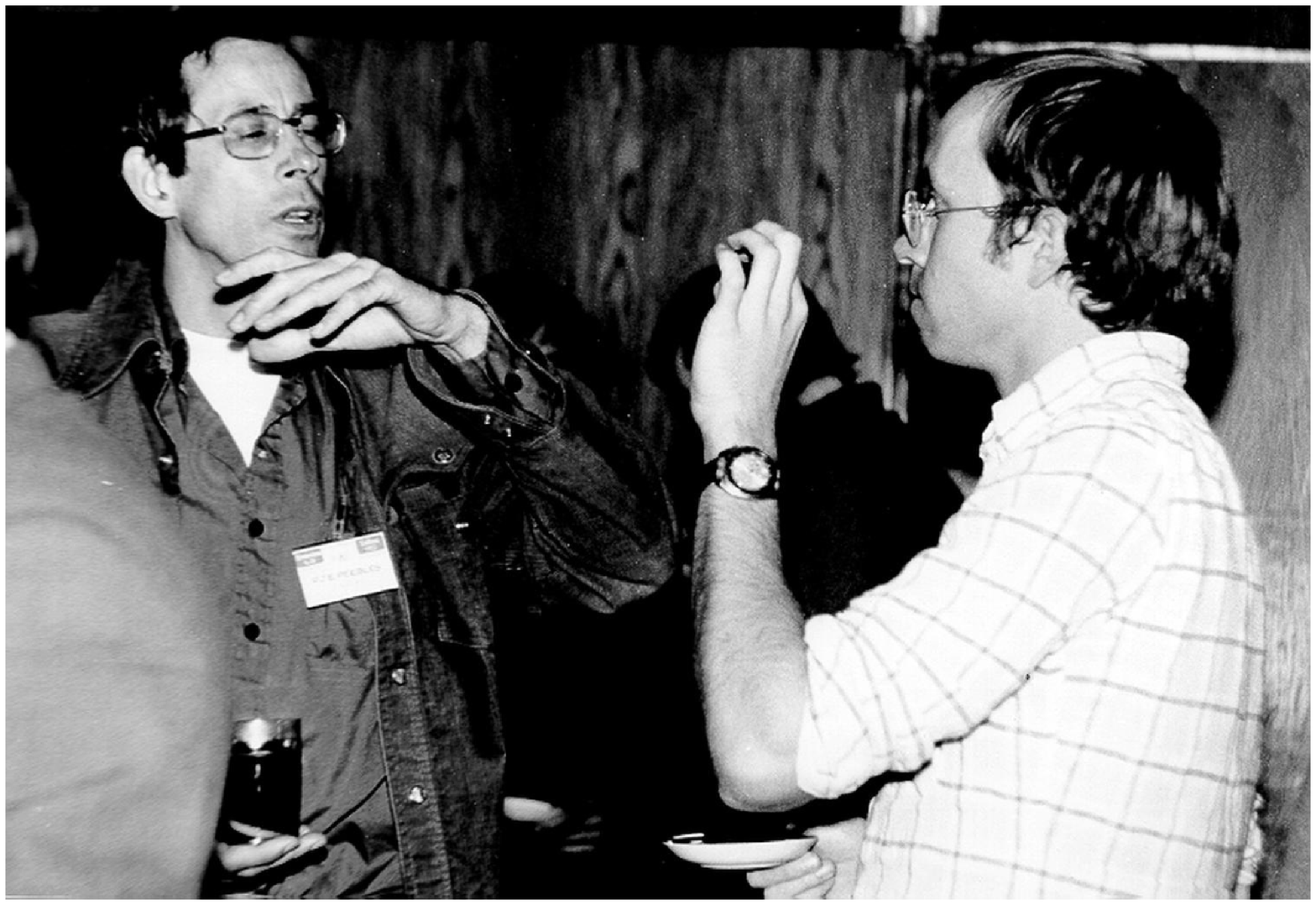}}
\resizebox{.40\columnwidth}{!}{\includegraphics*{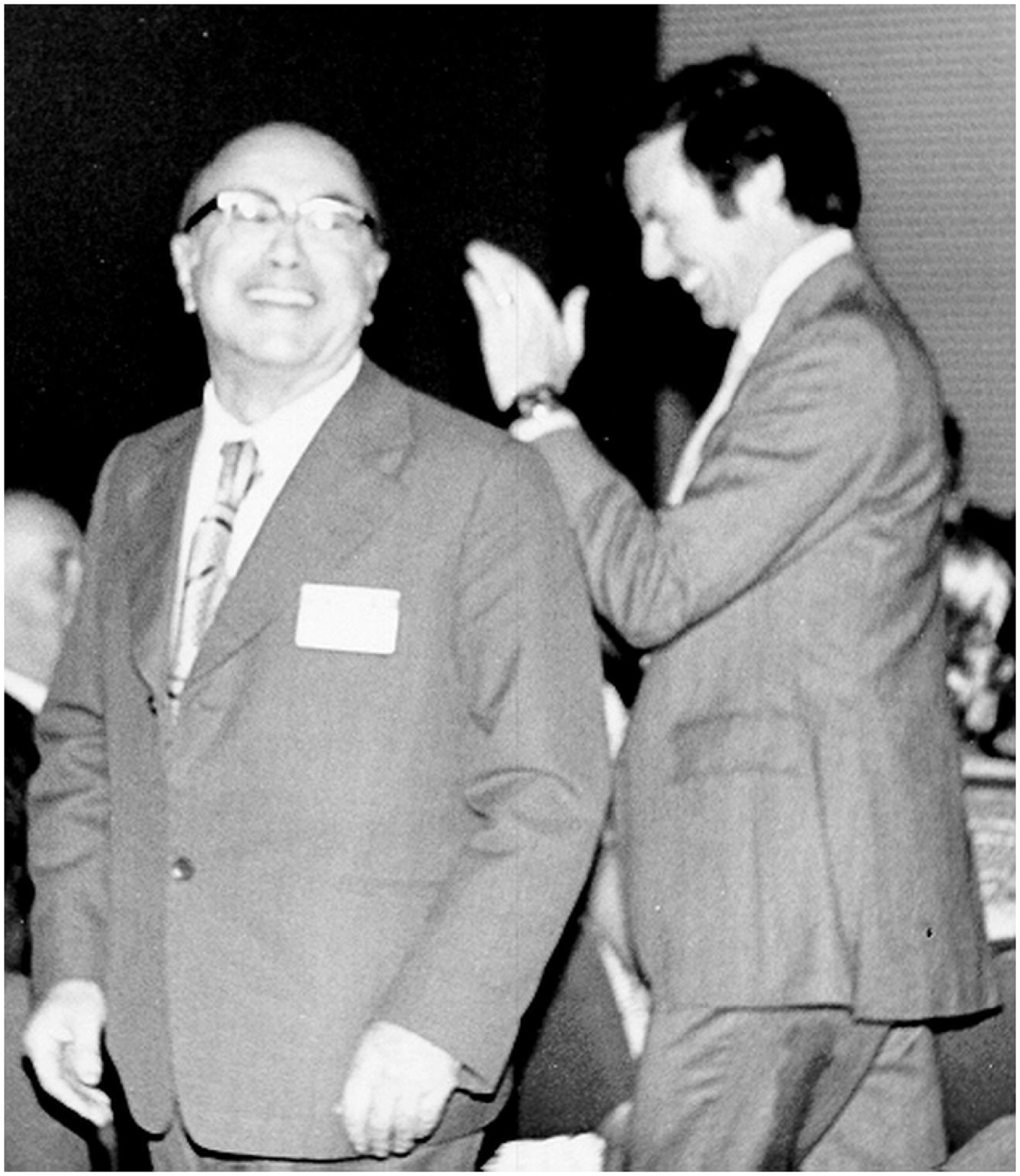}}
\caption{Participants of the Tallinn Symposium on the Large Scale
  Structure of the Universe. Left: Peebles and Tremaine;
  Right: Zeldovich and Longair. }
\label{fig:tallinn77}
\end{figure*}

\subsection{Understanding the large-scale structure of the Universe}

The first problem to solve was to find some explanation for the absence
of galaxies in voids.  This was done by Einasto, J\~oeveer \&
Saar\cite{Einasto:1980}.  Saar developed an approximate model of the
evolution of density perturbations in under-- and over--dense regions
based on Zeldovich ideas.  He found that the matter flows out of
under--dense regions and collects in over--dense regions until it
collapses (pancake forming) and forms here galaxies and clusters.  In
under--dense regions the density decreases continuously, but never
reaches zero: there must be primordial matter in voids.  In these under-dense
regions the density is always less than the mean density, thus galaxy formation is
not possible.

Initially we believed that pancakes are 2--dimensional surfaces as predicted
by Zeldovich\cite{Zeldovich:1970}.  To our surprise we did not find evidence
for the presence of wall--like structures between voids -- {\em the dominating
  structural element was a chain (filament) of galaxies and clusters}. 
The absence of wall--like pancakes and the dominance of filaments was
explained theoretically by Bond, Kofman \& Pogosyan\cite{Bond:1996}, through
the effect of tidal forces.

The presence of filaments and voids in the galaxy distribution was met with
some scepticism.  One objection against the new concept was raised by Peebles:
the human brain has a tendency to see regularity (filaments in galaxy
distribution) even in the case if the actual distribution is almost random.
Zeldovich\cite{Zeldovich:1978} addressed this problem in his talk during the
IAU Symposium and suggested that objective criteria should be used to check
this aspect of the galaxy distribution. Together with the Zeldovich team we
developed several methods to characterise the filamentary distribution. One of
these methods was the multiplicity function of galaxy systems, it is sensitive
to the richness of galaxy systems. The other test applied was the connectivity
or percolation test, which makes difference between galaxy systems consisting
of isolated clusters, clusters connected with filaments, and a random
distribution of galaxies. 

Applying these tests to real galaxy samples confirmed the presence of galaxy
systems of very different richness, and a high connectivity due to galaxy
filaments between superclusters. The application of tests to simulated
structures showed that the original Zeldovich scenario (where small-scale
perturbations were absent), as well as the Peebles scenario, have some
problems. The Zeldovich scenario passes the percolation test, but not the
multiplicity test -- there are no fine structures within superclusters. The
Peebles scenario fails both tests, results of these tests were described by
Zeldovich, Einasto \& Shandarin\cite{1982Natur.300..407Z} and Einasto et
al.\cite{1984MNRAS.206..529E}.

Difficulties with both scenarios were solved when the Cold Dark Matter model of structure formation was applied, which unites best properties of both the Zeldovich and the Peebles scenarios (Melott et al.\cite{Melott:1983}, Blumenthal et al.\cite{Blumenthal:1984}). Galaxies form by clustering and merging of smaller galaxies as suggested by Peebles, but their formation starts in future superclusters, where the density is highest, as it follows from the Zeldovich pancake scenario.

At the time of the Tallinn symposium only relatively small areas of sky were
covered by complete magnitude-limited galaxy samples, thus many astronomers
were suspicious to the presence of the overall cellular distribution. It was
evident that wide--area and deeper flux--limited galaxy redshift surveys are
needed.  Harvard astronomers made for the whole Northern hemisphere a survey
up to limiting magnitude 14.5, later the survey was extended to 15.5
magnitude, CfA1 and CfA2 surveys, respectively.  The second CfA surveys shows
the filamentary character of the galaxy distribution very clearly, as seen
from the first slice of their study by de Lapparent, Geller \&
Huchra\cite{de-Lapparent:1986}. The presence of the cosmic web was confirmed.

\begin{figure*}[ht]
\centering
\resizebox{.90\columnwidth}{!}{\includegraphics*{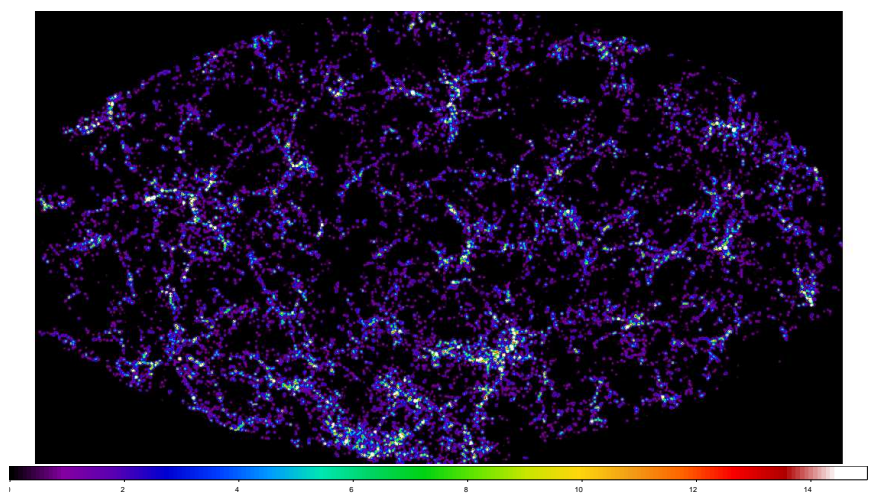}}
\caption{Sloan Digital Sky Survey, Northern section: a shell of thickness
  10~\Mpc\ at a distance $d=240$~\Mpc\ from us, as seen on sky in coordinates
  $x=-d ~\lambda$ and $y=d ~\eta \cos\lambda$, where $\lambda$ and $\eta$
  are intrinsic SDSS spherical coordinates. Filamentary
  superclusters of various richness, voids and weak filaments crossing voids
  are well seen. The rich complex in the lower region is the Great Wall, a complex
  of several very rich superclusters. }
\label{fig:sdss}
\end{figure*}

A much deeper redshift survey up to the blue magnitude 19.4 was made using the
Anglo-Australian 4-m telescope.  This Two degree Field Galaxy Redshift Survey
(2dFGRS) covers an equatorial strip in the Northern Galactic hemisphere and a
contiguous area in the Southern hemisphere\cite{Colless:2001}.  Over 250
thousand redshifts have been measured.  Presently the largest project to map
the Universe, the Sloan Digital Sky Survey (SDSS), has been completed by a
number of American, Japanese and European universities and
observatories\cite{2009ApJS..182..543A}.  The goal was to map a quarter of the
entire sky: to determine positions and photometric data in 5 spectral bands of
galaxies and quasars, and redshifts of all galaxies down to red magnitude {\tt
  r = 17.7} (about 1 million galaxies).  All 7 data releases have been made
public. This has allowed to map the largest volume of the Universe so
far. Fig.~\ref{fig:sdss} shows the luminosity density field of a shell of
thickness 10~\Mpc\ at a distance $d=240$~\Mpc\ from us.

The discovery of the non--baryonic nature of the Dark Matter has resolved
the first fundamental problem of the Dark Matter, discussed by Tammann
in Tbilisi in 1975.  The second problem, the smoothness of the Hubble
flow, was explained only recently with the discovery of Dark Energy,
previously called also cosmological lambda term.  Two teams, led by
Riess\cite{Riess:1998} (High-Z Supernova Search Team) and Perlmutter
\cite{Perlmutter:1999} (Supernova Cosmology Project), initiated
programs to detect distant type Ia supernovae in the early stage of
their evolution, and to investigate with large telescopes their
properties. These supernovae have an almost constant intrinsic
brightness. By comparing the luminosities and redshifts of nearby and
distant supernovae it is possible to calculate how fast the Universe
was expanding at different times. The supernova observations give
strong support to the cosmological model with the $\Lambda$ term.

Studies of the Hubble flow in the nearby space, using observations of
type Ia supernovae with the Hubble Space Telescope (HST), were carried
out by several groups.  The major goal of the study was to determine
the value of the Hubble constant.  As a by-product also the smoothness
of the Hubble flow was investigated.  One of these projects was led by Allan
Sandage\cite{Sandage:2006}.  The analysis confirmed earlier results that the
Hubble flow is very quiet over a range of scales from our Local
Supercluster to the most distant objects observed. This smoothness in
spite of the inhomogeneous local mass distribution requires a special
agent.  Sandage emphasises that no viable alternative to Dark Energy
is known at present, thus the quietness of the Hubble flow gives
strong support for the existence of Dark Energy.  This effect has been
investigated in detail by Arthur Chernin\cite{Chernin:2003a}.

\section{Conclusions}

\begin{itemize}

\item{} The presence of Dark Matter in the Universe and the Cosmic Web
  were established gradually by astronomers from many centers, in several
  cases Tartu astronomers have pioneered in these studies.

\item{} Initially the Dark Matter concept had many problems, until its
  non--baryonic nature and the presence of Dark Energy were found
  ($\Lambda$CDM model). Also the concept of the web-like distribution of
  galaxies was initially met with scepticism, which disappeared when wide-area
  and complete redshift surveys were completed.

\item{} The discoveries of Dark Matter and Cosmic Web are connected:
  Dark Matter as the dominant population in the Universe determines
  properties of the Web, and the structure of the Web gives
  information on properties of DM particles.

\item{} In the study of Dark Matter and the structure of the Universe our
  Tartu team benefited from the earlier experience in Tartu by \"Opik,
  Rootsm\"ae and Kuzmin, and especially from a close collaboration with the
  Moscow cosmology team led by Yakov Zeldovich.

\end{itemize}

\section*{Acknowledgements}

I thank Maret Einasto, Gert H\"utsi, Juhan Lauri Liivam\"agi, Enn Saar, Erik
Tago, Elmo Tempel, and our colleague Volker M\"uller in Potsdam for
collaboration and for the permission to use our common results in this review
talk.  My special thank is to late Mihkel J\~oeveer for pioneering
contributions both in the search for Dark Matter and of the structure of the
Universe, as well as to late Yakov Borissovich Zeldovich and his team, who
inspired us to study the large--scale distribution of galaxies, and
participated in the development of the ideas following the discovery of the
cosmic web.  The present study was supported by the Estonian Science
Foundation grant ETF 4695, and by grant TO 0060058S98.  I thank ICRAnet, and
the Astrophysikalisches Institut Potsdam (DFG-grant MU 1020/15-1) where
part of this study was performed, for hospitality.

\bibliographystyle{ws-procs975x65}

%\bibliography{mg12.bib}

\end{document}